\documentclass[apj]{emulateapj}

\usepackage{graphicx}
\usepackage{amsmath}
\usepackage{cases}

\usepackage{color}\def\red#1{\textcolor{black}{#1}}

\usepackage{natbib}

\newcommand{\msun}{\, M_\odot}
\newcommand{\pcc}{{\rm cm^{-3}}}

\def\bfrac#1#2{\left(\frac{#1}{#2}\right)}
\def\bmfrac#1#2{\left(-\frac{#1}{#2}\right)}
\def\vec#1{\mbox{\boldmath $#1$}}
\def\pd#1#2{\frac{\partial #1}{\partial #2}}

\shorttitle{R-process cosmic rays}
\shortauthors{Komiya}

\begin{document}
\title{R-process Element Cosmic Rays from Neutron Star Mergers}

\author{Yutaka Komiya\altaffilmark{1} and Toshikazu Shigeyama\altaffilmark{1}}
\altaffiltext{1}{Research Center for the Early Universe, University of Tokyo, Hongo 7-3-1, Bunkyo-ku, 113-0033, Tokyo, Japan}

\begin{abstract}
Neutron star mergers (NSMs) are one of the most plausible sources of r-process elements in the universe. 
Therefore NSMs can also be a major source of ultra-heavy elements in cosmic rays. 
In this paper, we first estimate the contribution of r-process elements synthesized in NSMs to the ultra-heavy element cosmic rays (UHCRs) by calculating transport equations that take into account energy loss processes and spallations. 
We show that the flux of UHCRs accelerated by NSMs themselves fluctuates by many orders of magnitude on the timescale of several million years
 and can overwhelm UHCRs accelerated by supernova remnants (SNRs) after an NSM takes place within a few kilo-parsec from the solar system. 
Experiments with very long exposure times using meteorites as UHCR detectors can detect this fluctuation. 
 As a consequence, we show that if NSMs are the primary source of UHCRs, future experiments using meteorites have a possibility to reveal the event history of NSMs in the solar vicinity. 
We also describe a possible difference in the abundance pattern and energy spectrum of UHCRs between NSM and SNR accelerations. 
\end{abstract}

\section{Introduction}

Coalescence of a close binary with two neutron stars or with a neutron star and a black hole are paid a great attention as 
 promising sources of gravitational wave and neutrino emission, and 
 plausible progenitors of short gamma-ray bursts (SGRBs) and other electromagnetic transients. 
Neutron star mergers (NSMs) are thought to be the most promising site for the rapid neutron capture process (r-process) nucleosynthesis,
 and a possible dominant source of r-process elements in the universe \citep{Lattimer74}. 
In this paper, we show that NSMs are also a possible major source of the ultra-heavy element component of cosmic rays.

The r-process is one of the major nucleosynthetic processes to synthesize elements heavier than the iron group \citep{B2FH}. 
The astronomical source of the r-process elements is a longstanding problem in nuclear astrophysics. 
Two scenarios have been proposed for the dominant astronomical sources of r-process elements. 
One is the core-collapse supernova (CCSN) scenario and the other is the NSM scenario \citep[e.g., review by][]{Cowan91}. 
In this paper, the term ``NSM" includes the coalescence of a neutron star - black hole binary.

Though CCSNe were widely accepted as a major source for r-process elements more than a decade ago, 
  there have been growing evidence supporting the NSM scenario in recent years. 
Theoretical studies of nucleosynthesis have revealed that elements at or above the second r-process peak are hard to be synthesized in the CCSNe \citep[e.g.][]{Wanajo11, Wanajo13}  
 excepting the model of the magneto-rotational driven explosion \citep[e.g.][]{Nishimura15}.
On the other hand, all of the NSM ejecta should become r-process elements because of their very low electron fractions. 
One NSM event yields r-process elements with a mass of $M_{r, \rm NSM} \sim 0.0001 - 0.1 \msun$ with a very high velocity of $v \sim 0.2 c$, where $c$ is the speed of light \citep[e.g.][]{Hotokezaka13}. 
Resent numerical studies of nucleosynthesis in NSMs successfully reproduce the solar r-process abundance pattern \citep[e.g.,][]{Bauswein13, Wanajo14}. 
The radioactive decay of these r-process nuclei powers a electromagnetic transient known as a kilonova or macronova, which emit photons mainly in the infra-red band. 
The observed infrared excess in the afterglow of GRB 130603B is consistent with kilonova models and supports NSMs as a major r-process source \citep{Tanvir13, Berger13, Barnes13, Tanaka13}.

NSMs have a low event rate, ${\cal R}_{\rm NSM} \sim 10^{-5}$/yr in the Milky Way (MW), while 
 the expected r-process yield is much larger than that of a CCSN ($M_{r, \rm SN} \sim 10^{-5} \msun$). 
As a result, both of the scenarios predict a similar production rate of r-process elements on average over a long period of time. 
From a viewpoint of chemical evolution, it has been argued that the rarity of the NSM results in much larger abundance scatter for metal-poor stars than in observations \citep{Argast04, Komiya14}. 
However, recent chemo-dynamical simulations with mixing of metal or semi-analytic modeling with large scale spreading of NSM ejecta predict the abundance scatter compatible with observations \citep{vandeVoort15, Hirai15, Shen15, Komiya16}. 
Furthermore, there are arguments using a radioactive r-process element $^{244}$Pu that indicate a low frequency of r-process production events. 
The abundance of $^{244}$Pu in the current interstellar medium (ISM) estimated from deep sea measurement is significantly lower than the early solar system abundance \citep{Wallner15}. 
\citet{Hotokezaka15} argued that this may indicate a very low event rate of the r-process sources. 
\citet{Tsujimoto17} investigated the chemical evolution of $^{244}$Pu in the solar vicinity and showed that the event frequency of r-process production is $\sim 1/1400$ of CCSNe.

Based on these studies, NSMs are thought to be a dominant source of r-process elements. 
If so, there can be a significant contribution from NSMs to r-process element nuclei also in cosmic rays because of very high velocities of the NSM ejecta. 
In this paper, we investigate cosmic ray originating from NSMs using a propagation model of cosmic ray   
 and discuss constraints on the contribution of NSMs by results of cosmic ray experiments.

Heavy elements in cosmic rays have been interested as a useful tool to investigate the source of the Galactic cosmic rays. 
It is known that the metal abundance relative to hydrogen and helium in cosmic rays are enhanced in comparison with these ratios in the solar atmosphere. 
This enhancement is thought to arise from acceleration mechanisms depending on the first-ionization-potential (FIP) and/or on the volatility and mass of an element. 
\citet{Meyer97} argue that material locked in grains is accelerated to cosmic ray energies more efficiently than ions in a gas phase.

Until recently, elements heavier than the iron group in cosmic rays (ultra-heavy element cosmic ray, hereafter UHCR) have been investigated in the context of the biased acceleration models by taking into account r-process elements supplied only from SNe. 
\citet{Binns89} reported the abundances of elements with $Z \geq 33$ measured by the {\it HEAO-3} satellite. 
The results show that, for $33 \leq Z \leq 60$, the cosmic ray source has a composition similar to that of the solar system when we consider the acceleration bias by FIP. 
The {\it SuperTIGER} instrument measured abundances of elements with $Z = 26$ - 40 in cosmic rays \citep{Murphy16}. 
They obtained results consistent with a model of cosmic ray origin with a source mixture of $19\%$ material from massive stars (stellar wind + supernova ejecta) and $81\%$ ISM material with the volatility biased preferential acceleration. 
On the other hand, the abundance pattern of elements with $Z > 60$ shows enhancement of r-process elements \citep{Binns89}.  
The measured UHCR composition shows the third peak of r-process abundance around Pt but do not show the third s-process peak at Pb. 
Later observations confirm the existence (absence) of the peak around Pt (Pb) \citep{Donnelly12, Alexeev16}. 
The r-process enhanced abundance pattern may indicate enhancement of r-process elements at the acceleration region of the cosmic ray,
 i.e., acceleration by the reverse shock in massive star material, though we should note here that the volatility bias model also predicts a large abundance ratio of Pt/Pb. 
The {\it Arel-6} satellite found overabundances of elements at $Z =$ 60 - 82 by a factor of $1.84 \pm 0.14$ \citep{Fowler87}.

Recently, \citet{Kyutoku16} investigated r-process elements in cosmic rays supplied from NSMs and argued that NSMs should enhance the UHCR flux by a few orders of magnitude compared to the solar composition. 
According to their study, the energy injection rate, $\dot{E}_r$, of r-process cosmic rays at the reverse shock of NSMs is described as follows (neglecting the acceleration bias), 
\begin{equation}
\dot{E}_{r, \rm NSM} = \frac{M_{r, \rm NSM}}{M_{\rm ej, NSM}} \epsilon_{\rm CR, RS} E_{\rm exp, NSM} {\cal R}_{\rm NSM} , 
\end{equation}
 where $M_{\rm ej, NSM}$ and $E_{\rm exp, NSM}$ are the mass and the kinetic energy ejected by a single NSM event, respectively,  $\epsilon_{\rm CR, RS}$ is the fraction of the ejecta kinetic energy to be converted to the cosmic ray energy at the reverse shock. 
In the case of NSMs, their ejecta are composed of pure r-process elements, i.e., $M_{r, \rm NSM}/M_{\rm ej, NSM} \sim 1$. 
The energy injection rate to UHCR through the reverse shock, $\dot{E}_{r, \rm SN}$, and forward shock, $\dot{E}_{r, \rm ISM}$, of SNe are described in a similar way, 
\begin{equation}
\dot{E}_{r, \rm SN} = \frac{M_{r, \rm SN}}{M_{\rm ej, SN}} \epsilon_{\rm CR, RS} E_{\rm exp, SN} {\cal R}_{\rm SN} . 
\end{equation}
\begin{equation}
\dot{E}_{r, \rm ISM} = X_{r, \rm ISM} \epsilon_{\rm CR, FS}  E_{\rm exp, SN} {\cal R}_{\rm SN} ,
\end{equation}
 where  $M_{\rm ej, SN}$ and $E_{\rm exp, SN}$ are the mass and kinetic energy of SN ejecta, respectively, 
 $X_{r, \rm ISM}$ is the abundance of the r-process elements in ISM, 
 $\epsilon_{\rm CR, FS}$ is the energy conversion rate to cosmic rays at forward shock, 
 and ${\cal R}_{\rm SN}$ is the event rate of SNe. 
\red{The contribution from the forward shock of NSMs to $\dot{E}_{r, \rm ISM}$ is negligible since ${\cal R}_{\rm NSM} \ll {\cal R}_{\rm SN}$ and $E_{\rm exp, NSM} \lesssim E_{\rm exp, SN}$. }
They thought the observed composition similar to the solar system indicates that 
 $\dot{E}_{r, \rm ISM}$ is larger  than one third of the other two components, 
 i.e., $3 \dot{E}_{r, \rm ISM} \gtrsim \dot{E}_{r, \rm SN}$ and $3 \dot{E}_{r, \rm ISM} \gtrsim \dot{E}_{r, \rm NSM}$. 
Since the r-process element abundance is $X_{r, \rm ISM} \sim 10^{-7}$ in ISM and ${M_{r, \rm SN}}/{M_{\rm ej, SN}} \sim 10^{-5}$ in CCSN ejecta, 
 the former constraint requires $\epsilon_{\rm CR, RS}/\epsilon_{\rm CR, FS} \lesssim 0.03$ in the CCSN scenario. 
Here, as mentioned above, the total r-process injection rate from NSMs is comparable to that from CCSNe, i.e., 
 $M_{r, \rm NSM} {\cal R}_{\rm NSM} \sim M_{r, \rm SN} {\cal R}_{\rm SN}$, 
while $E_{\rm exp}/M_{\rm ej}$ of an NSM is one hundred times or more larger than that of a CCSN because of the large velocity of NSM ejecta. 
Therefore, 
$3\dot{E}_{r, \rm ISM} \gtrsim \dot{E}_{r, \rm NSM}$ indicates 
$\epsilon_{\rm CR, RS}/\epsilon_{\rm CR, FS}<0.0003$. 
They concluded either that NSM is not the main origin of r-process elements or that the
 acceleration of cosmic rays is very inefficient at the reverse shock in NSM ejecta.

Here it should be noted that their study does not consider energy loss and decay processes of cosmic rays in ISM. 
UHCRs are significantly affected by spallation by collision with particles in ISM \citep{Waddington96, Combet05}.
Furthermore, NSM is so rare that the long time interval between NSM events in the solar neighborhood enables a significant decay of cosmic rays from NSMs.   
In this paper, we estimate the UHCR flux from NSMs taking into account energy loss processes and spallation by collision with nuclei in ISM.

The kinematics of particles ejected from an NSM is poorly understood. 
There is no numerical study to investigate how heavy ions dissipate their energy through collisionless shocks. 
From observational sides,
 emission from shock heated material at the heads of jets from micro-quasars with a velocity of $\sim 0.2 c$ have been observed \citep[e.g.][]{Dubner98, Diaz Trigo13}.  
These can be interpreted as dissipation at a collisionless shock but 
 the radiation is predominantly emitted from electrons 
 and the energy dissipation rate for baryonic component is unknown. 
Thus we consider two scenarios about their kinematics and production of UHCR in this paper.

In one scenario, the NSM ejecta propagate in a similar way to SN ejecta which forms a forward shock in the ISM and a reverse shock in the ejecta \citep{Montes16}.  
In this scenario, 
 the kinetic energy of NSM ejecta is dissipated through the reverse shock and used to push the ISM.  
Most of the ejecta particles lose their kinetic energy while some particles become cosmic rays through the diffusive shock acceleration after a reverse shock forms in the ejecta.

The other scenario is the propagation of the NSM ejecta without shock accelerations. 
\citet{Tsujimoto14} pointed out that
 the NSM ejecta cannot be treated as a fluid because of their very large stopping lengths 
 and the ion component of ejecta can propagate into a very large volume.  
For example, the kinetic energy of elements moving at a speed of $0.2 c$ is 19 MeV/nucleon and the stopping length of $^{153}$Eu with this energy is 2.6 kpc \citep[see][]{Komiya16}.

From the perspective of the Galactic chemical evolution of r-process elements, 
\citet{Komiya16} showed that the abundance distribution of extremely metal-poor stars in the Milky Way halo is well reproduced under the NSM scenario only when we adopt the large scale spreading of the NSM ejecta. 
If ions in the NSM ejecta propagate into the interstellar space with velocities similar to the initial velocities of the NSM ejecta as they argued, 
 these high energy ions can be observed as ``cosmic rays'' even without an additional acceleration in the reverse shock.

This paper is organized as follows. 
In Section~\ref{timescaleS}, we assess the timescales of decay processes and propagation of UHCR. 
In Section~\ref{modelS}, we describe our model to compute the NSM origin cosmic ray flux. 
In Section~\ref{resultS}, we present results of our computations. 
We conclude the paper in Section~\ref{conclusionS}.

\section{Decay and Propagation}\label{timescaleS}
\begin{figure}
\includegraphics[width=\columnwidth]{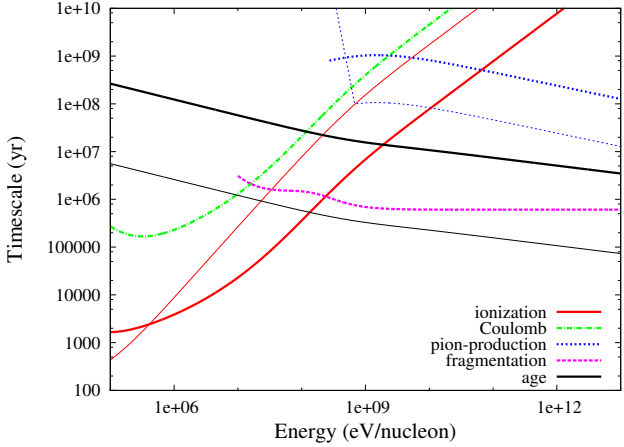}
\caption{
Timescales of the ionization energy loss (red solid), 
Coulomb scattering (green dash-dotted), 
pion pair-creation (blue dotted), 
and spallation (magenta dashed)
as a function of energy per nucleon. 
Thick and thin lines represent values for xenon, as a representative of r-process elements, and proton, respectively. 
The thick black line show the propagation timescale, which gives the typical age of a last NSM event which affects r-process cosmic rays. 
The thin black line shows the propagation timescale of cosmic ray protons from SNRs. 
}\label{timescale}
\end{figure}

Cosmic ray nuclei lose their energy in ISM by ionization of neutral hydrogens, Coulomb scattering of electrons, and pion pair creation. 
In addition, heavy element cosmic ray collides with ISM particles and fragment into lighter elements. 
We describe these decay processes in this section.  
Figure~\ref{timescale} summarizes their timescales and compares with the propagation timescale (age) of UHCRs.

\subsection{Energy Loss of Cosmic Ray Nuclei due to Interaction with ISM}
We use the formulae for the energy loss rate of cosmic ray nuclei due to interactions with ISM by \citet{Schlickeiser02}.

The ionization energy loss rate is described using the following formula, 
\begin{eqnarray}
\dot{E}_{\rm ion} &=& -1.82 \times 10^{-7} Z_{\rm eff}^2 \bfrac{n_{\rm H}}{1 \pcc} \nonumber  \\ 
&\times& (1+0.0185 \log \beta H(\beta-\beta_0)) \frac{2\beta^2}{\beta_0^3+2\beta^3} \ {\rm eV/s} , 
\end{eqnarray}
 where $Z_{\rm eff}$ is the effective charge, 
 $n_{\rm H}$ is the number density of the neutral hydrogen in ISM, 
 $\beta = v/c$,  $H$ the Heaviside step function, 
 and $\beta_0( = 0.01)$ is the orbital velocity of electrons in hydrogen atoms.  We assume $n_{\rm H} = 1 \pcc$ in the following.
The effective charge is approximated as $Z_{\rm eff} = Z (1 - 1.034 \exp(-137\beta Z^{-0.688}))$, 
 which is less than the charge $Z$ of the nucleus, since ultra-heavy ions are not completely stripped at small energies. 

The energy loss rate by Coulomb scattering is approximated as 
\begin{equation}
\dot{E}_{\rm Coulomb} = -3.1\times 10^{-7} Z_{\rm eff} \bfrac{n_{\rm e}}{1 \pcc} \frac{\beta^2}{2.34\times10^{-5}+\beta^2} \ {\rm eV/s}
\end{equation}
where $n_{\rm e}$ is the number density of electrons in ISM and we set $n_{\rm e} = 0.01 \pcc$ \red{\citep{Wolfire95}. }

The energy loss rate by pion production is described as follows, 
\begin{equation}
\dot{E}_{\rm pp} = -4.9 c m_\pi c^2 n_{\rm H} H(\gamma-1.3) \sigma_{\rm pp} \gamma ,  
\end{equation}
 where $m_\pi$ is the mass of pion and 
 $\gamma$ is the Lorentz factor of a cosmic ray particle. 
$\sigma_{\rm pp}$ is the pion creation cross section and approximated as 
\begin{eqnarray}
\sigma_{\rm pp} = \begin{cases}
 6.13 \times 10^{-26} \bfrac{E_{\rm k}}{m_{\rm p}c^2}^{7.64} \gamma^{-0.25} \ {\rm cm^2} & \left( \frac{E_{\rm k}}{m_{\rm p}c^2} < 0.75 \right)  \\
 8.12 \times 10^{-27} \bfrac{E_{\rm k}}{m_{\rm p}c^2}^{0.53} \gamma^{-0.25} \ {\rm cm^2}  & \left( \frac{E_{\rm k}}{m_{\rm p}c^2} \geq 0.75 \right) 
\end{cases}
\end{eqnarray}
 where $E_{\rm k}$ is the kinetic energy of the particle and $m_{\rm p}$ is the proton mass.

\subsection{Spallation}
UHCR nuclei can fragment into lower mass nuclei in inelastic collisions with atoms of the interstellar gas. 
We use an empirical formula for the total cross-section of spallation by \citet{Letaw83}. 
The cross-section at high energies ($> 2 {\rm GeV}/A$) is independent of energy, 
\begin{equation}
\sigma_{\rm f, HE} = 45 A^{0.7} (1+0.0016\sin(5.3-2.63\log A)) \ {\rm mb} , 
\end{equation}
 where $A$ is the mass number. 
At energy below $2 {\rm GeV}/A$, the cross-section is given as a function of energy and mass number as, 
\begin{eqnarray}
\sigma_{\rm f, LE} = \sigma_{\rm f, HE} && \left[ 1 -  0.62 \exp \left( -\frac{E_{\rm k}/A}{200 {\rm MeV}}\right) \right. \nonumber \\
&& \left. \sin \left( 10.9 \bfrac{E_{\rm k}/A}{{\rm MeV}}^{-0.28} \right) \right] .
\end{eqnarray}
This formula is valid for $E_{\rm k}/A>$10 MeV.  
Below this energy, we neglect spallation since the energy loss due to ionization dominates fragmentation loss.

\subsection{\red{Propagation} of the Cosmic Ray Nuclei}
When cosmic rays propagate following the diffusion equation, 
 the timescale for propagation over a distance $r$ is given as 
\begin{equation}\label{eq:diftime}
t = \frac{r^2}{D}, 
\end{equation}
where $D$ is the diffusion coefficient.  
On the other hand, the event rate of NSMs taking place in a cylindrical region of the MW disk with a radius $r$ around the sun is estimated to be ${\cal R}_{\rm NSM} (r/r_{\rm MW})^2$, where $r_{\rm MW}$ is the radius of the MW disk.  
Therefore, the typical timescale from the last NSM event in this region is 
\begin{equation}\label{eq:eventtime}
t = {\cal R}_{\rm NSM}^{-1} \bfrac{r_{\rm MW}}{r}^2 .
\end{equation} 
From equation~(\ref{eq:diftime}) and (\ref{eq:eventtime}), we obtain the propagation timescale, 
\begin{equation}
t_{\rm CR} = r_{\rm MW} (D {\cal R}_{\rm NSM})^{-1/2},
\end{equation} 
 \red{which gives the typical age of a last NSM (or SNR) event which affects UHCRs at the solar system. 
When the decay timescale is smaller than $t_{\rm CR}$,
 we have to consider the decay processes (spallation and/or energy loss) to estimate the UHCR flux. }

\subsection{Comparison of Timescales} 

In Figure~\ref{timescale}, we plot the energy loss timescale, $|E/\dot{E}|$, the fragmentation loss timescale, $(n_{\rm H} v \sigma_{\rm f})^{-1}$, and the propagation timescale, $t_{\rm CR}$, of cosmic rays. 
As a representative of r-process elements, we show the values of xenon, the second peak element of the r-process. 
We also plot the timescales for proton with thin lines for comparison.

Nuclei with energies $\lesssim$100 MeV/$A$ lose their energy predominantly due to ionization.  
The energy loss timescale of an ultra-heavy element is $\sim 10^5$ yrs at 20 MeV/$A$, $\sim 1$ dex shorter than that of a proton because of their larger electric charges. 
The energy loss rate by Coulomb scattering is much smaller than that due to ionizations as far as $n_{\rm e} \ll n_{\rm H}$. 
The rate of pion creation is very small and negligible in this study. 
In the energy range above $ \sim 100\,{\rm MeV}/A$, 
 spallation is the dominant decay process of r-process elements, 
 and its timescale is about one million years.

The solid black lines in Figure~\ref{timescale} show the propagation timescales (minimum ages) of UHCRs from NSMs (thick) and protons from SNRs (thin), 
 where we set $r_{\rm MW} = 20$ kpc, ${\cal R}_{\rm NSM} = 10^{-5}\, {\rm yr^{-1}}$ and ${\cal R}_{\rm SN} = 3 \times 10^{-2}\, {\rm yr^{-1}}$. 
We adopt the diffusion coefficient given in \citet{Thoudam14}, 
\begin{equation}\label{eq:D}
D(\rho) = 5 \times 10^{28} \beta \bfrac{\rho}{3{\rm GV}}^{0.33}\, {\rm cm^2 s^{-1}} ,
\end{equation} 
 where $\rho  = {pc}/{Ze}$ is the particle rigidity 
 and $p$ is the momentum of a nucleus and $e$ is the elementary charge.

As shown in the figure, the propagation timescale of cosmic rays from NSMs is $\sim 10^8$ yr at 1 MeV/$A$ and $\sim 10^7$ yr at 1 GeV/$A$, 
 and longer than the decay timescale by ionization or spallation. 
In particular, energy loss timescale at $v \sim 0.2 c$ (19 \,MeV/$A$) is $\sim 3$ dex shorter than the propagation timescale. 
This indicates that we have to consider energy loss and fragmentation to estimate the cosmic ray flux from NSMs.
On the other hand,  the energy loss timescale of protons from SNRs is about ten times longer than that of r-process elements and thus the energy loss is not important for cosmic ray protons accelerated in SNRs excepting at $\lesssim 20$ MeV/$A$. 
In addition, we should not assume equilibrium between cosmic ray input and energy loss, and 
 have to treat each NSM event discretely since the time interval between NSM events in the solar vicinity is longer than the energy loss timescale.

\section{Model Description}\label{modelS}

We introduce a numerical model to compute the propagation of cosmic rays in the MW considering the processes discussed in the previous section.

\subsection{Input of Cosmic Rays}

We compute four models for the source term of UHCRs.  
We assume that all the r-process elements are synthesized only in NSMs. 
These r-process nuclei are assumed to be accelerated to UHCRs in NSMs (referred to as NSM-UHCRs) or in SNRs (SNR-UHCRs). 
We consider three models for the acceleration in NSMs to compensate the poor understanding of the kinematics of the ejecta particles as pointed out in Introduction.

The first model is the UHCRs accelerated by the reverse shock in the NSM ejecta (model NSM-RS).  
If the NSM ejecta propagate as a blast wave like a usual SNR, 
 a part of the ejecta particles is accelerated to UHCRs via diffusive shock acceleration  
 while most of the particles are trapped in the ejecta shell. 
Though a forward shock of the blast wave can accelerate r-process nuclei in the ISM, the forward shock component is negligible because $X_{r, \rm ISM} \ll M_{r, \rm NSM}/M_{\rm ej, NSM}$ and the much lower event rate of NSMs than that of SNe. 

The second and third models assume unshocked ejecta from NSMs. 
As mentioned in Introduction, nuclei in NSM ejecta have possibility to spread into ISM with large kinetic energy without shock acceleration and have energies of MeV/$A$ or higher. 
Since the energy spectrum of these unshocked nuclei is not well known, we test two energy spectra; 
one is the Maxwell-Boltzmann distribution (model NSM-MB) and the other is \red{a} power law distribution \red{with the index of -2} (model NSM-P).

We also consider the UHCRs accelerated by the diffusive shock acceleration in SNRs as considered in previous studies (model SNR). 
In this model, r-process elements in ISM having the solar abundances (i.e., $X_{r, \rm ISM}=X_{r_\odot}$) are accelerated at forward shocks by SNRs. 
\red{ In the case of SNRs, we take into account the acceleration bias. }
We neglect r-process elements synthesized by a SN explosion and accelerated by the reverse shock. 
We compute the fluxes and spectra of SNR-UHCRs and NSM-UHCRs to compare them with observations. 
We also compute the flux of cosmic ray protons and irons accelerated in SNRs in order to calibrate our model 
 and to give r/Fe ratio in cosmic rays.

Parameters for the input of cosmic rays in each model are determined as follows.

\subsubsection{Event Rate}

In our model, each NSM or SN event occurs discretely with a finite time interval. 
NSMs and SNe occur randomly at rates proportional to the surface density of the MW disk.

The event rate of NSMs and SNe in the MW is set to be 
$10^{-5} {\rm yr^{-1}}$ and
$3\times 10^{-2} {\rm yr^{-1}}$, respectively. 
They are assumed to take place on the equatorial plane of the disk ($z=0$). 
For the radial density profile of the disk, we adopt the exponential disk model, $\propto \exp(-r/r_s)$, with the scale radius $r_s = 3$ kpc and the size of 20 kpc.

\begin{figure}
\includegraphics[width=\columnwidth]{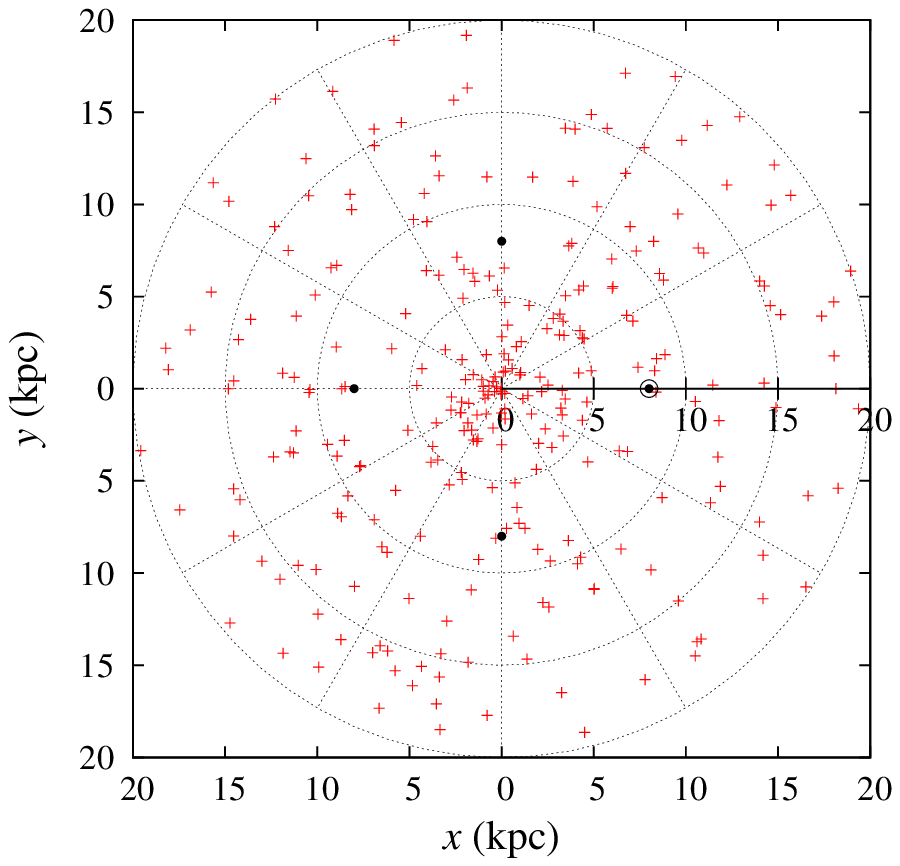}
\includegraphics[width=\columnwidth]{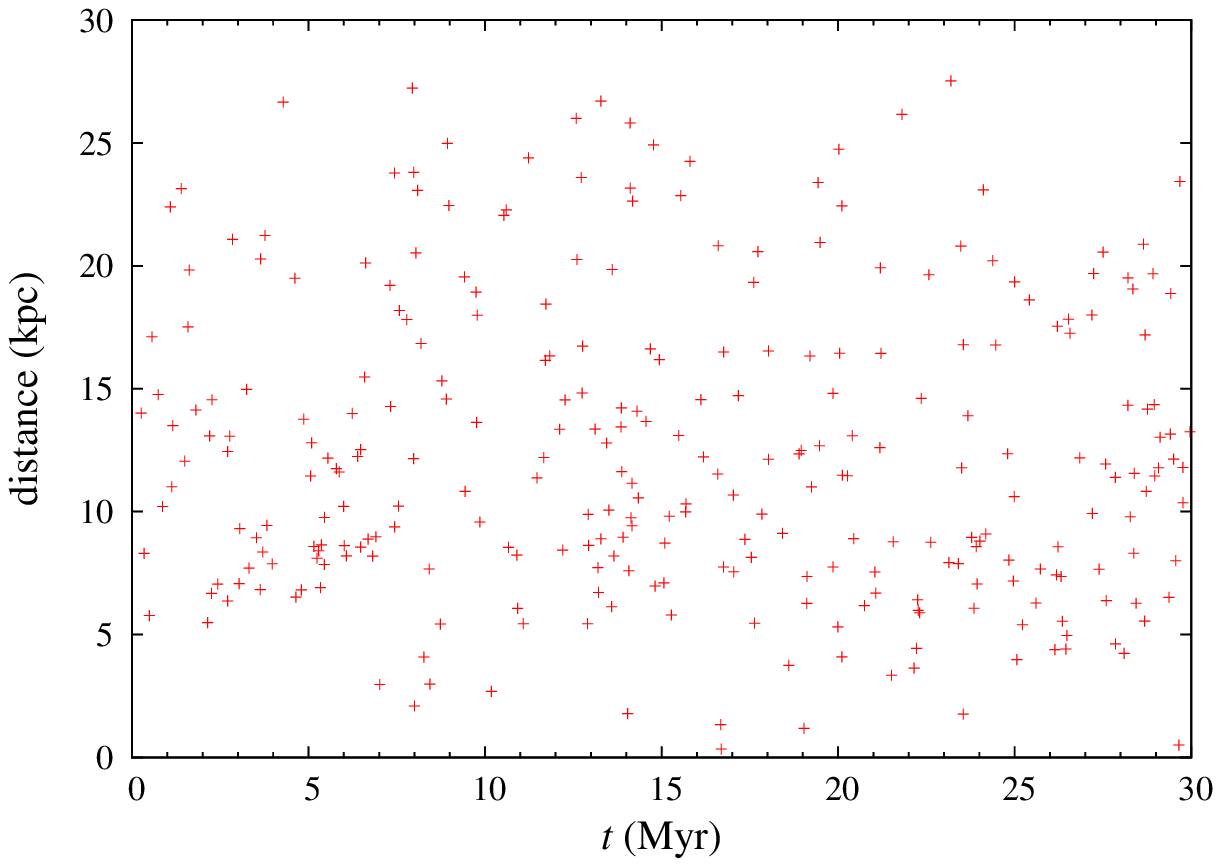}
\caption{
Top panel: Positions of NSM events from $t=0$ to $3\times 10^7$ yr in our model. 
The black circle is the position of the solar system. 
Bottom panel: Time of occurrence of NSM events and their distances from the solar system. 
}\label{events}
\end{figure}

We plot the positions and time points of occurrence of NSMs in our computation in Figure~\ref{events}. 
In this figure, we only plot events in the period from $t = 0$ through $3\times 10^7$ yr though the computation covers the time span between  $t = -3\times 10^7$ yr and $t = 2 \times 10^8$ yr.

\red{
The mass of r-process elements produced by a single event is set to be $M_{\rm ej, NSM} = 0.01\msun$
 so that NSMs with the event rate of $10^{-5} {\rm yr^{-1}}$ can account for all the r-process elements in the Milky Way assuming all stars have roughly solar abundances. 
}

\subsubsection{Shock Accelerated Cosmic Rays}

The source energy spectra for models SNR and NSM-RS are assumed to follow a power-law distribution with respect to momentum
\begin{equation}
Q(p) \propto p^{-q} , 
\end{equation}
We set the index $q=2.3$
 and the minimum kinetic energy to be 300 MeV$/A$.

The proportionality constant is determined from the total energy $ \int Q(p) E_{\rm k}(p) dp = E_{\rm exp} \epsilon_{\rm CR} X_r b$, 
 where $b$ is a parameter to describe the acceleration bias. 
The explosion energy of a SN is set to be $ E_{\rm exp, SN} = 10^{51}$ erg. 
For an NSM, we set $ E_{\rm exp, NSM} = (\gamma_{\rm ej}-1) M_{\rm ej, NSM} c^2$,  
 where $\gamma_{\rm ej}$ is the Lorentz factor of the NSM ejecta of $v = 0.2c$. 
The energy conversion efficiency, $\epsilon_{\rm CR}$, to cosmic rays has a large uncertainty. 
In this paper, we show a result with $\epsilon_{\rm CR, FS} = 0.1$ for the forward shock and 
$\epsilon_{\rm CR, RS} = 0.03$ for the reverse shock. 
$X_r = 10^{-7}$ for model SNR and $X_r = 1$ for the NSM ejecta. 
We use \red{$X_{\rm Fe} = 0.0013$} and $X_{\rm H} = 0.7$, for iron and hydrogen, respectively.

In this study, we introduce a bias parameter $b$ in model SNR.  
As mentioned in Introduction, the acceleration efficiency of heavy elements is thought to be higher than protons. 
In addition, there is the possible contribution of nuclei accelerated at reverse shocks in SN ejecta. 
Since both these effects are not well understood, we determine the value of $b$ to reproduce the observed cosmic ray flux of iron. 
\red{ As shown later, the model with $b = 200$ well reproduces the observations. }
We use the same $b$ for r-process elements for simplicity. 
In model NSM-RS, the biased acceleration should not work ($b = 1$) since most of the cosmic ray nuclei in NSM ejecta are ultra-heavy elements.

The computed UHCR flux is simply proportional to the energy input $E_{\rm exp} \epsilon_{\rm CR} X_r b$ to UHCRs by one event. 
In model SNR, $E_{\rm exp} \epsilon_{\rm CR} X_r b = 2\times 10^{45}$ erg and the number of UHCR particles produced by one event is $N_r = 4.4 \times 10^{45}$. 
In the NSM-RS model, $E_{\rm exp} \epsilon_{\rm CR} X_r b = 1.1\times 10^{49}$ erg and $N_r = 2.4\times 10^{49}$.

\subsubsection{Unshocked NSM Ejecta}

We also consider the unshocked NSM-UHCRs in this study. 
The velocity distribution of NSM ejecta has been investigated by numerical simulations for coalescence of a NS binary. 
\citet{Nagakura14} fitted the density profile of homologously expanding NSM ejecta in the simulation of \citet{Hotokezaka13} by a power law distribution $\rho \propto r^{-n}$, 
 and yielded $n = 3 - 4$. 
Since the homologous expansion indicates that $r=vt$, the velocity distribution of the ejecta also follow the power law, 
\begin{equation}\label{eq:Qv2}
Q(v) \propto v^{-n+2} .
\end{equation}
The high energy tail of the velocity distribution still has significant uncertainty. 
\citet{Hotokezaka13} show that the typical maximum velocity of the ejected material is $0.5 - 0.8c$. 
The velocity distribution of the ejecta presented by \citet{Radice16} shows exponential decay at high energy. 
On the other hand, observations of SGRBs may indicate the existence of very high energy ($\gamma \sim 100 - 1000$) component in the NSM ejecta, though it might be composed of pure leptons.

Thus we adopt the following two energy spectra. 
One is the power law in terms of velocity (model NSM-P) to mimic the result of \citet{Hotokezaka13}.   
When we write the distribution of equation~(\ref{eq:Qv2}) as a function of momentum, 
\begin{equation}
Q(p) \propto  \gamma(p)^{n-5} (m_p A)^{n-3} p^{-n+2} .
\end{equation}
Here, we use $n=4$. 
The other is the relativistic Maxwell-Boltzmann distribution (model NSM-MB), 
\begin{equation}
Q(p)dp \propto p^2 \exp \bmfrac{\sqrt{A^2 m_{\rm p}^2 c^4 + p^2 c^2}}{kT} dp ,   
\end{equation}
which reproduces the exponential cutoff at high energies obtained in \citet{Radice16}.

The low velocity cutoff in model NSM-P is determined so that the average energy of the ejecta particle is $(\gamma_{\rm ej}-1) A m_{\rm p} c^2$. 
The cutoff energy is $E_{\rm min,P} = 0.2$ MeV/$A$ in model NSM-P. 
$kT$ in model NSM-MB is also set to yield the same average energy\red{, and $kT/A = 13 {\rm MeV}/A$}. 
The proportionality constant is determined to satisfy $ A m_{\rm p} \int Q(p) dq = M_{\rm ej, NSM}$ and $M_{\rm ej, NSM} = 0.01 \msun$.   
In these two models, 
 the total kinetic energy is $E_{\rm exp, NSM} = 3.7\times 10^{50}$ erg and the particle number is $N_r = 9.2 \times 10^{52}$.

Thick lines in the top panel of Figure~\ref{evolve} show the input energy spectra for models NSM-RS (blue solid), NSM-MB (red dashed), and NSM-P (green dotted). 
Model NSM-MB shows a steep decay above $\sim 100$ MeV/$A$. 
Model NSM-P has the largest input flux at $\sim 1$ GeV/$A$ while model NSM-RS have an enhanced component at very high energies above $\sim1$ PeV/$A$.

\subsection{Diffusion\red{, Fragmentation, and Energy Loss}}

The propagation of cosmic rays is governed by the diffusion equation, 
\begin{equation}
\pd{N(E, \vec{r}, t)}{t} = \nabla (D \nabla N) - \pd{(N \dot{E})}{E} -N n_{\rm H} v \sigma_{\rm f} +Q(E,\vec{r},t) ,
\end{equation}
 where $N$ is the number density of cosmic ray nuclei, the energy loss rate $\dot{E}$ is the sum of the three processes described in \S 2, $\dot{E} = \dot{E}_{\rm ion} + \dot{E}_{\rm Coulomb} + \dot{E}_{\rm pion}$, and $Q(E,\vec{r},t)$ is the source term.

When cosmic rays with an energy of $E_0$ is injected at a position $\vec{r}_0$ and time $t_0$, i.e., $Q(E,\vec{r},t) = \delta(E-E_0) \delta(\vec{r}-\vec{r}_0) \delta(t-t_0)$, 
 the solution of the diffusion equation is described by the Green function,  
\begin{equation}\label{eq:Green}
 G(E,E_0, \vec{r},\vec{r}_0, t,t_0) = 
  \frac{\exp(-\nu)}{(4\pi \lambda)^{3/2}} \exp \bmfrac{(r-r_0)^2}{4 \lambda} \delta(E_0-E-\epsilon) . 
\end{equation}
Here $\epsilon$ is the energy loss of a particle, 
\begin{equation}\label{eq:epsilon}
\epsilon(E_0, \tau) = \int^{\tau}_{0} \dot{E} dt, 
\end{equation}
and $E(E_0, \tau) = E_0 + \epsilon(E_0, \tau) $ is the energy of a particle with an initial energy $E_0$ at time $\tau=t-t_0$, 
$\nu$ is the interaction depth weighted by the spallation cross section,  
\begin{equation}\label{eq:nu}
\nu(E_0, \tau) = \int^{\tau}_{0} n_{\rm H} v(E(E_0, t)) \sigma_{\rm f}(E(E_0, t))dt , 
\end{equation}
and $\lambda^{1/2}$ is the propagation length defined as 
\begin{equation}\label{eq:lambda}
\lambda(E_0, \tau) = \int^{\tau}_{0} D(E(E_0, t)) dt ,
\end{equation}
 where $D(E)$ is the diffusion coefficient as a function of the particle energy given in equation~(\ref{eq:D}).

\begin{figure}
\includegraphics[width=\columnwidth]{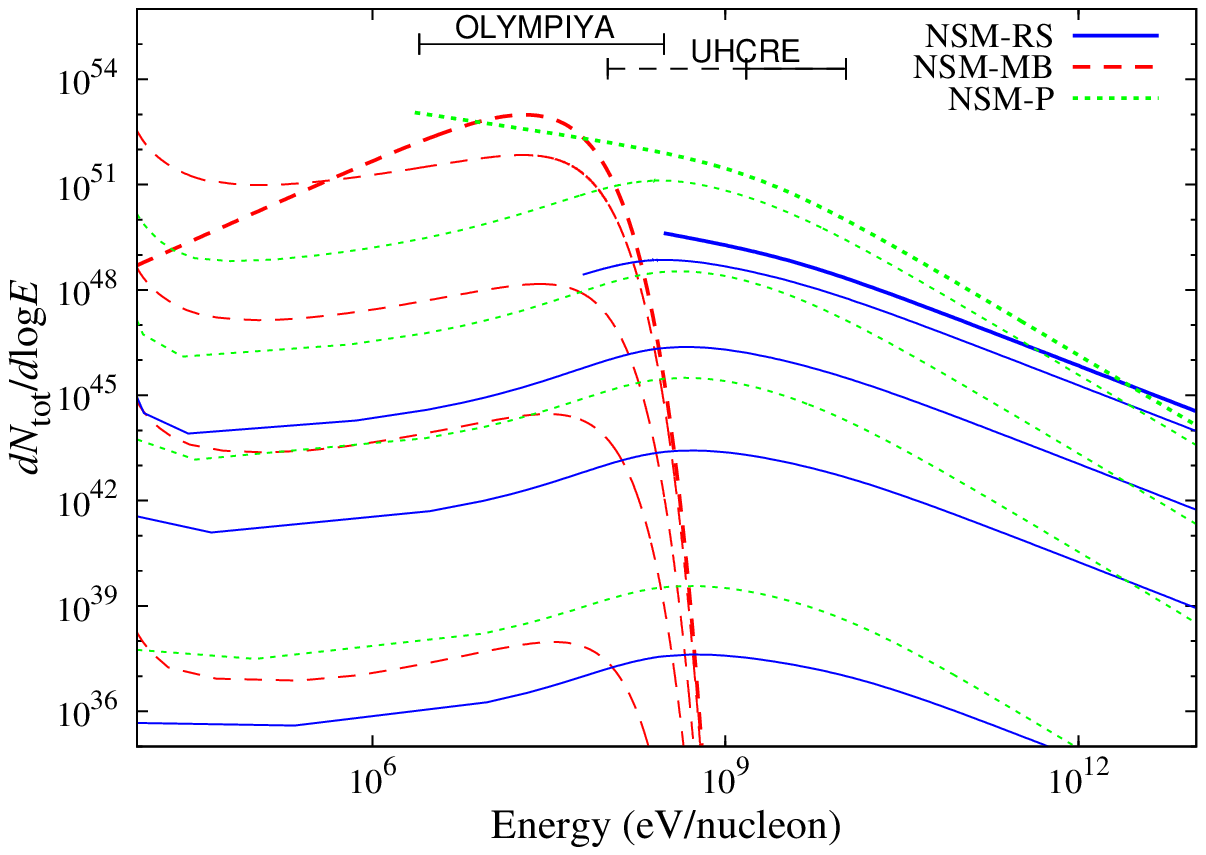}
\includegraphics[width=\columnwidth]{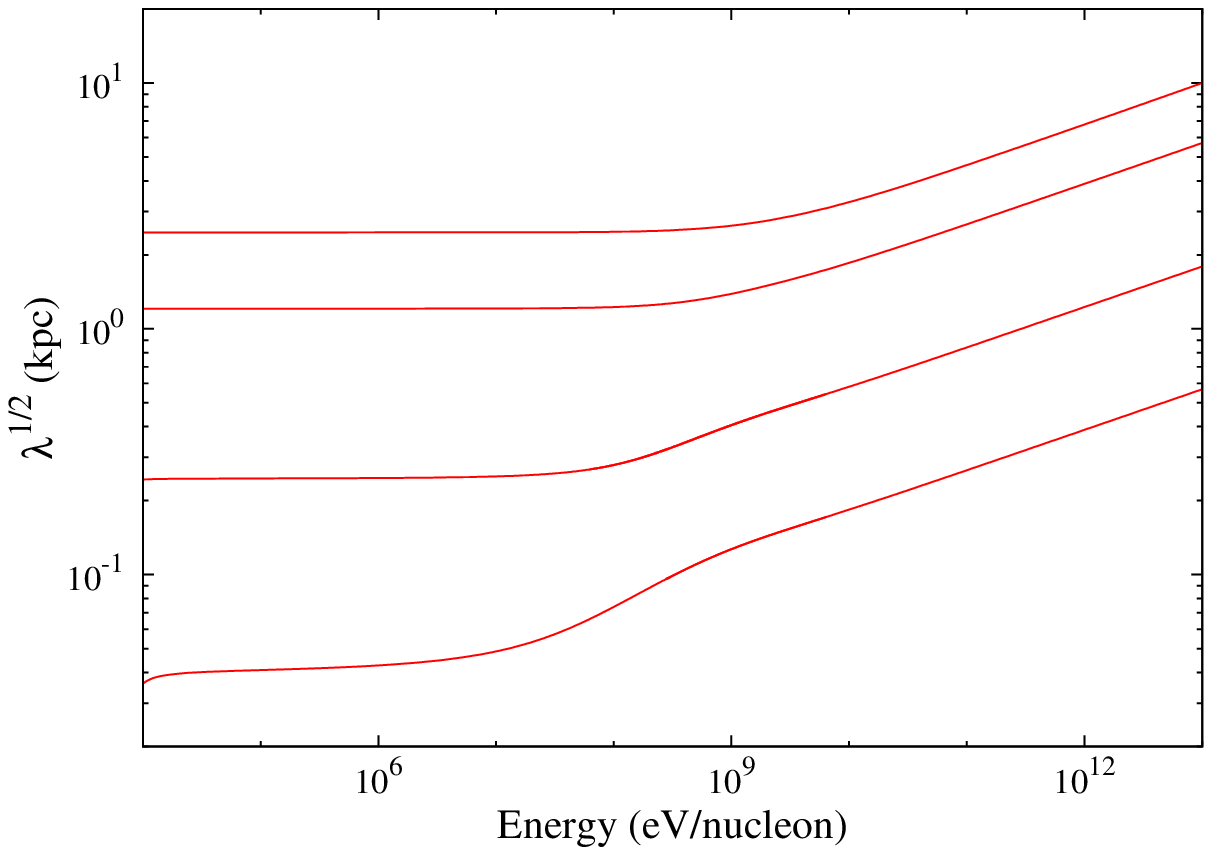}
\caption{
Top panel:
Temporal evolution of the energy spectra of particles from an NSM. 
\red{ For model NSM-RS  (blue solid lines) and  NSM-P  (green dotted lines), we plot spectra at $t=0, 10^6, 5\times10^6, 10^7,$ and $2\times10^7$yr, from top to bottom. 
For models NSM-MB  (red dashed lines), spectra are plotted for a shorter time span, i.e., $t=0, 10^5, 5\times10^5, 10^6,$ and $2\times10^6$yr. }
Horizontal bars at the top indicate the detectable energy range of an instrument on the UHCRE satellite and stony iron meteorites. 
The detectors of UHCRE can detect UHCRs in the energy range of the dashed bar while
 the geomagnetic cutoff energy for the satellites is at 1.5 GeV/$A$. 
Bottom panel: 
The diffusion distance, $\lambda^{1/2}$, as a function of energy $E(t)$. 
The values at $t=10^5, 10^6, 10^7,$ and $3\times10^7$ yr are plotted from bottom to top. 
}\label{evolve}
\end{figure}

Since cosmic rays under consideration are injected from multiple events with the input energy spectrum of $Q(E_0)$, 
 the number density of cosmic rays is obtained by integrating over the source energy and summing the contribution from all events,  
\begin{equation}\label{eqn:output}
N(E,\vec{r},t) = \sum_i \int Q(E_0)  G(E, E_0, \vec{r}, \vec{r}_i, t, t_i) dE_0 , 
\end{equation}
 where $r_i$ and $t_i$ is the position and time point of occurrence of the $i$-th event.

In this study, we calculate the evolution of each cosmic ray particle with an energy in one of equidistant logarithmic energy bins according to equations~(\ref{eq:epsilon}) (\ref{eq:nu}), and (\ref{eq:lambda}). 
Then we obtain the energy spectrum of cosmic rays by a convolution of the resultant particle distribution with the input energy spectrum as in equation (\ref{eqn:output}). 
The input energy spectrum is covered by the 14,000 energy bins ranging from 300 MeV/$A$ to 3 PeV/$A$ for models NSM-RS. 
For model NSM-P, we use 20,000 energy bins in the range of $ E_{\rm min,P}$ - $10^{10} \times E_{\rm min,P} $ 
 and for model NSM-MB, 12,000 energy bins covering the range of $5\times10^{-4} kT $ - $ 5\times10^2 kT$. 
Figure~\ref{evolve} shows the evolution of comic rays from a single NSM event which occurs at $t_0=0$. 
The top panel is the energy spectra, $ dN_{\rm tot}/d\log E$, 
 where $ N_{\rm tot}$ is the total number, $ N_{\rm tot}(E, t) = \iiint \int^E N(E, \vec{r},t) dEd\vec{r} $, of UHCR nuclei with energy below $E$, 
 and the bottom panel shows the propagation length $\lambda^{1/2}$.

\begin{figure}
\includegraphics[width=\columnwidth]{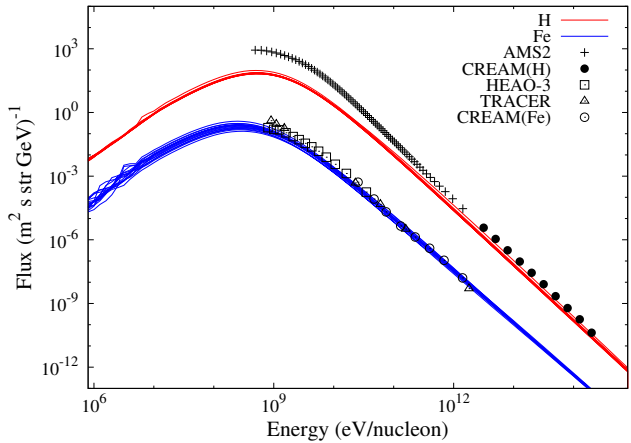}
\caption{
The predicted energy spectra of the cosmic ray protons (red) and iron (blue) at the solar system. 
\red{ We overplot 15 lines at different time points at $2 \times 10^5$ yr intervals. }
Crosses and filled circles show the observational results of protons in AMS2 \citep{Aguilar15} and CREAM \citep{Yoon11}, respectively. 
Squares, triangles, and open circles are the observed flux of iron by HEAO-3 \citep{Engelmann90}, TRACER \citep{Obermeier11} and CREAM \citep{Ahn09}, respectively. 
}\label{proton}
\end{figure}

Figure~\ref{proton} shows the resultant cosmic ray energy spectra for protons and iron at the solar system in our model. 
As shown in the figure, 
 our model almost reproduces the observed flux and energy spectra of the cosmic ray protons and iron. 
Here, we adopt $b=200$ for iron. 
The observed flux of protons is slightly higher than our model result \red{at $\lesssim 10^{12}$eV}. 
\citet{Thoudam14} showed that the observed enhancement is explained by the re-acceleration of cosmic rays in diffused SNR shock. 
In their result, the re-acceleration is negligible for iron or heavier elements while important for protons. 
In this study, we neglect the re-acceleration since we focus on ultra-heavy elements.

\red{
\subsection{Observations}
We observe the flux of UHCRs at present using satellites and balloons. 
The average flux of  UHCRs over the past millions of years can be obtained using meteolites. 
Pallasites, a kind of stony iron meteorite, can be used as a detector of UHCRs. 
UHCRs create tracks in olivine crystal in meteorites due to induced structure transformations and broken bonds and the meteorite preserve the tracks for millions of years. 
}

\red{
In this paper, we estimate the UHCR flux detected by the {\it Ultra-Heavy Cosmic-Ray Experiment (UHCRE)} on the {\it Long Duration Exposure Facility (LDEF)} satellite \citep{Donnelly12}, 
 and the stony iron meteorites. 
}

\red{
\subsubsection{Solar Modulation}
The Galactic cosmic ray flux at a low energy range is affected by the magnetic field coupled to the solar wind. 
The solar modulation can be described by the following formula, 
\begin{equation}
N_{\rm obs}(E,t) = N(E+\Phi_r, \vec{r}_\odot, t) \frac{E (E+2 T_r)}{(E+\Phi_r)(E+\Phi_r+2 T_r)} 
\end{equation}
where $T_r = 0.938 {\rm GeV}/A$, $\Phi_r = \frac{e Z}{A} \phi$ \citep{Usoskin11},  
 and $\vec{r}_\odot$ is the position of the solar system and $|\vec{r}_\odot| = 8$ kpc. 
 The modulation potential, $\phi$, is dependent on the solar activity. 
}
 
\red{
The UHCRE experiment measured the UHCRs from 1984 April to 1990 January. 
The average modulation potential at the the observation period is $\phi = 680$ GV \citep{Usoskin11}. 
This value is similar to the average of $\phi$ over 70 years. 
In the cases of meteorites, the solar modulation depends on the activity of the sun at when the meteorites stay in the interplanetary space, and the orbits of the meteorites. 
To investigate the solar modulation in the past millions of years is beyond the scope of this study. 
For simplicity, we use the same value, $\phi = 680$ GV, for meteorite observations. 
}

\subsubsection{Detection rate}

We compute the cosmic ray flux that can be detected by current observations,
\begin{equation}
F(t) = \int^{E_{\rm upp}}_{E_{\rm low}}  \frac{\red{N_{\rm obs}(E,t)} \beta}{4\pi} dE ,
\end{equation}
 where $E_{\rm upp}$ and $E_{\rm low}$ are the upper and lower threshold to be detected by instruments.

The detector of UHCRE is calibrated to detect ultra-heavy elements ($Z\geq 70$) in an energy range from 100 MeV/$A$ to 10.6 GeV/$A$ \citep{Donnelly12}. 
The geomagnetic cutoff for UHCRs at the orbit of LDEF was $\sim1.5$ GeV$/A$.  
Other satellite experiments also observe particles in the similar energy range. 
Here, we set $E_{\rm low, sat} = 1.5$ GeV$/A$ and $E_{\rm upp, sat} = 10.6$ GeV$/A$ for the satellite experiment.


In the {\it OLIMPIYA} project, \citet{Alexeev16} measured the abundance pattern of UHCRs using two stony iron meteorites as UHCR detectors. 
Most of the measured tracks were produced by nuclei with $Z \geq 50$. 
The calibration experiment is performed using particles with energies of 2.5 - 11.1 MeV$/A$.  
The higher energy threshold to create measurable tracks in meteorites is $\sim 300$ MeV$/A$ for Pb. 
Here, we set $E_{\rm low, met} = 2.5$ MeV$/A$ and $E_{\rm upp, met} = 300$ MeV$/A$ for the meteorite experiments. 

In the top panel of Figure~\ref{evolve}, we show the detectable energy ranges of these experiments with horizontal bars.

\section{Results and Discussion}\label{resultS}

\subsection{Flux}

\begin{figure}
\includegraphics[width=\columnwidth]{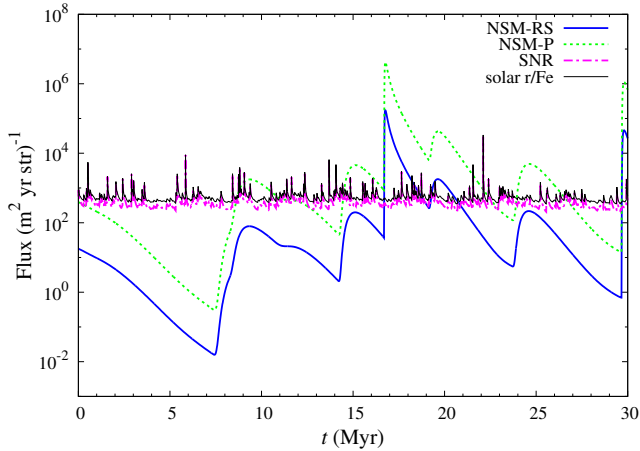}
\caption{\red{The predicted UHCR flux in the detectable energy range by the UHCRE satellite as a function of time. 
The solid blue and dotted green lines denote results of model NSM-RS and NSM-P, respectively. 
The dash-dotted magenta line denotes SNR-UHCRs. 
The black line shows the predicted flux of iron in the SNR model times $(X_r/X_{\rm Fe})_\odot$. 
In this energy range, the result of NSM-MB is far smaller and not plotted. }
}\label{satellite}
\end{figure}

\begin{figure}
\includegraphics[width=\columnwidth]{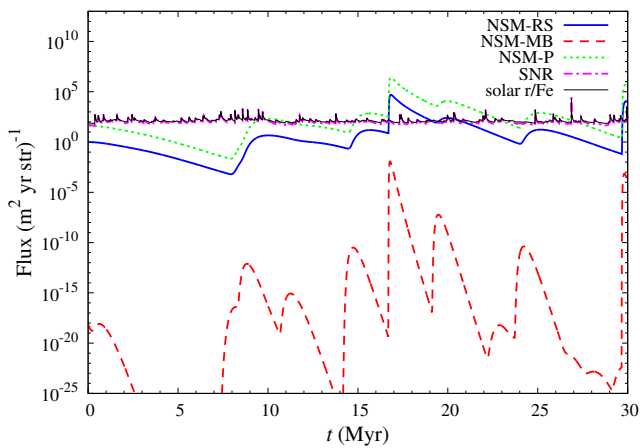}
\caption{\red{The predicted UHCR flux in the detectable energy range by meteorites. 
The red dashed line show the result of model NSM-MB. 
Line types and colors of other lines are the same as Figure~\ref{meteo}. }
}\label{meteo}
\end{figure}

We show the temporal evolution of intensities of UHCRs detectable by the satellite experiment and the pallasite measurements described in the previous section in \red{Figures~\ref{satellite} and \ref{meteo}.  
Figure~\ref{satellite} shows the flux in the energy range of UHCRE on a satellite, and
Figure~\ref{meteo} shows the flux of UHCRs in the energy range detectable by stony-iron meteorites.  }
The UHCR flux from NSMs fluctuate wildly on a timescale of millions of years 
 because of a very low event rate of NSMs and high rates of energy loss and spallation of UHCRs.

\red{
\subsubsection{Model NSM-RS}
The predicted flux fluctuates with many orders of magnitude over the period of a few Myr. 
An NSM eject many UHCRs but the NSM-UHCR flux decreases by collisional fragmentation, as discussed in section~\ref{timescaleS}.  
In the energy range for satellite experiments, spallation has a shorter timescale and dominates the decay rate of the flux. 
Though the meteorite experiments detect nuclei with 2.5 - 300 MeV, these nuclei had energies around 1 or 2 GeV/$A$ when they were ejected from an NSM several million years ago. 
Therefore,  the flux fluctuates also due to spallation. 
}

\red{
The predicted peak flux detectable by satellites becomes higher than that of SNR-UHCRs when an NSM takes place at $\lesssim 1.5$ kpc from the solar system. 
In this computation run, the flux of model NSM-RS overwhelms SNR-UHCR for $8\%$ of the simulated time. 
The value is dependent on the parameter $E_{\rm exp} \epsilon_{\rm CR} X_r b$. 
A ten times smaller bias $b=20$ in model SNR would increase the percentage to $22\%$. 
}


\red{
\subsubsection{Model NSM-MB}
The unshocked ejecta from NSMs with the Maxwell-Boltzmann energy distribution (NSM-MB) has a negligible contribution to the satellite experiments
 since the most of particles in the ejecta have energies lower than the geomagnetic cutoff at the orbit of the satellites. 
}

\red{
Also in the meteorite energy range, 
 the predicted flux is much smaller than SNR-UHCR even shortly after a NSM event at solar vicinity, 
 since the solar wind reduce the NSM-UHCR flux. 
By the solar modulation, meteorites can detect only cosmic rays with initial energy above $E_{\rm low, met} + \Phi_r$. 
In the NSM-MB model, the flux above $E_{\rm low, met} + \Phi_r$ is much lower, and the detected flux reduced by $\sim 10^{-6}$ by the solar modulation. 
In the NSM-RS model, on the other hand, the contribution of the solar modulation is much weaker. 
The flux from $E_{\rm low, met} + \Phi_r$ to $E_{\rm up, met} + \Phi_r$ is higher than at $E_{\rm low, met}$ - $E_{\rm up, met}$, and the flux in the meteorite energy range is only reduced by $\sim 1/4$. 
In the satellite energy range, the modulation amplitude is $\sim20\%$. 
}

\red{
The amplitude of temporal fluctuation of the flux is much larger than model NSM-RS. 
This is because the energy loss rate in ISM becomes high at low energy range. 
The energy loss timescale around the peak energy of NSM-MB is two or three orders of magnitude shorter than the propagation timescale. 
Model NSM-RS yields nuclei with energies of $\sim 1$ GeV, which come into the meteorite energy range by losing their energies due to ionization energy loss on timescales of Myrs
 and alleviate the decrease of the flux, 
 but model NSM-MB do not have such a high energy nuclei. 
}

\red{
\subsubsection {Model NSM-P}
Model NSM-P predicts 20 - 40 times higher flux than model NSM-RS in both meteorite and satellite energy ranges
 since all the nuclei becomes cosmic rays in this model. 
The typical flux of UHCR nuclei is comparable to model SNR. 
}

\red{
The decay curve resembles that of model NSM-RS and determined by spallation, 
 because both models have similar slopes of the spectra in the satellite energy range. 
Also for the meteorites, the slope above $E_{\rm low, met} + \Phi_r$ is similar. 
}




\subsubsection{Model SNR}
The flux of SNR-UHCRs also shows a fluctuation by 1 dex or more when a SN takes place in the solar vicinity. 
The flux of cosmic ray iron times $(X_r/X_{\rm Fe})_\odot$ in the SNR model is plotted with black lines. 
The cosmic ray r/Fe ratio in model SNR is almost same with the solar value since SNRs accelerate particles in ISM with the solar abundance. 

\subsubsection{Observations}
Observational data of HEAO-3 and Ariel-6 show that
 the abundance of r-process elements relative to iron is similar to the solar system abundance ratio. 
When the predicted flux of NSM-UHCR is similar to the black lines in Figures~\ref{satellite} and \ref{meteo}, our result is consistent with the observed r/Fe ratio. 
As shown in Figs.~\ref{satellite} and \ref{meteo}, all models can be consistent with observations in some time intervals. 
\red{The chance probability that the NSM-UHCR flux relative to iron from SNRs become comparable (1/3 to 3) to the solar abundance ratio is $15\%$ in model NSM-RS. }

Our results are in contrast to the constraint, $\epsilon_{\rm CR, RS}/\epsilon_{\rm CR, FS} < 0.0003$, by \citet{Kyutoku16}. 
The predicted flux from NSMs in our model can be smaller than from SNRs even when we assume $\epsilon_{\rm CR, RS} = 0.03$. 
This difference is due to spallation and the energy loss by ionization, and difference in the acceleration bias which they did not consider.

When the NSM-UHCR flux is much smaller than SNR-UHCR, NSM-UHCRs is obscured and it is difficult to find observational signatures of NSM-UHCRs by satellite observations. 
However, meteorite experiments can find the signatures of NSM-UHCRs as discussed in the next subsection.

In this study, the decay rate by spallation is thought to be overestimated because we neglect secondary cosmic ray particles formed by this process. 
Some spallation products can also be detected as UHCRs 
 and alleviate the drop of the total UHCR flux. 
On the other hand, ejecta mass from NSMs may have significant diversity, though we assume each NSM ejects the same amount of r-process elements. 
\citet{Hotokezaka13b} showed that a neutron star - black hole merger can eject about ten times larger masses than a double neutron star merger. 
The variety of the ejecta mass can enhance the amplitude of fluctuation of the UHCR flux.

\subsection{Cumulative Flux}

\begin{figure*}
\includegraphics[width=0.5\textwidth]{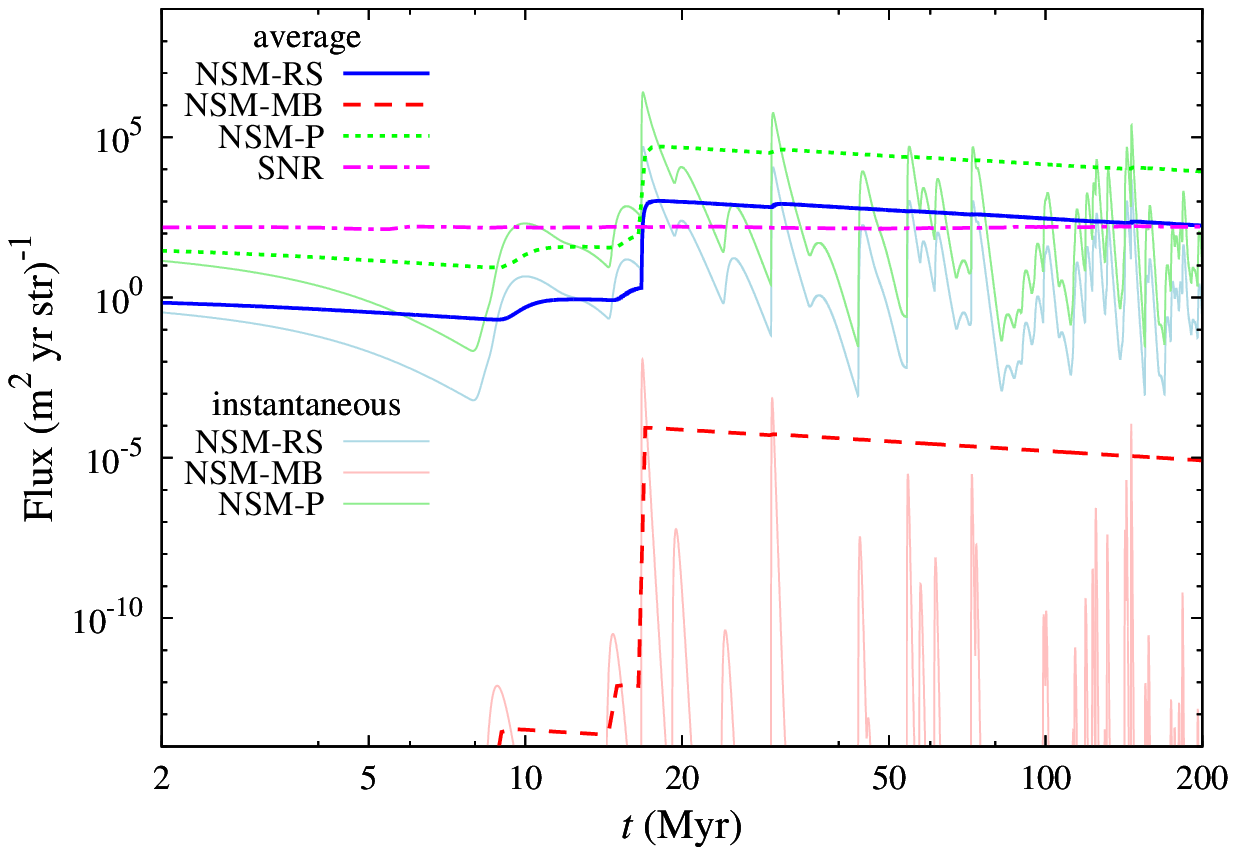}
\includegraphics[width=0.5\textwidth]{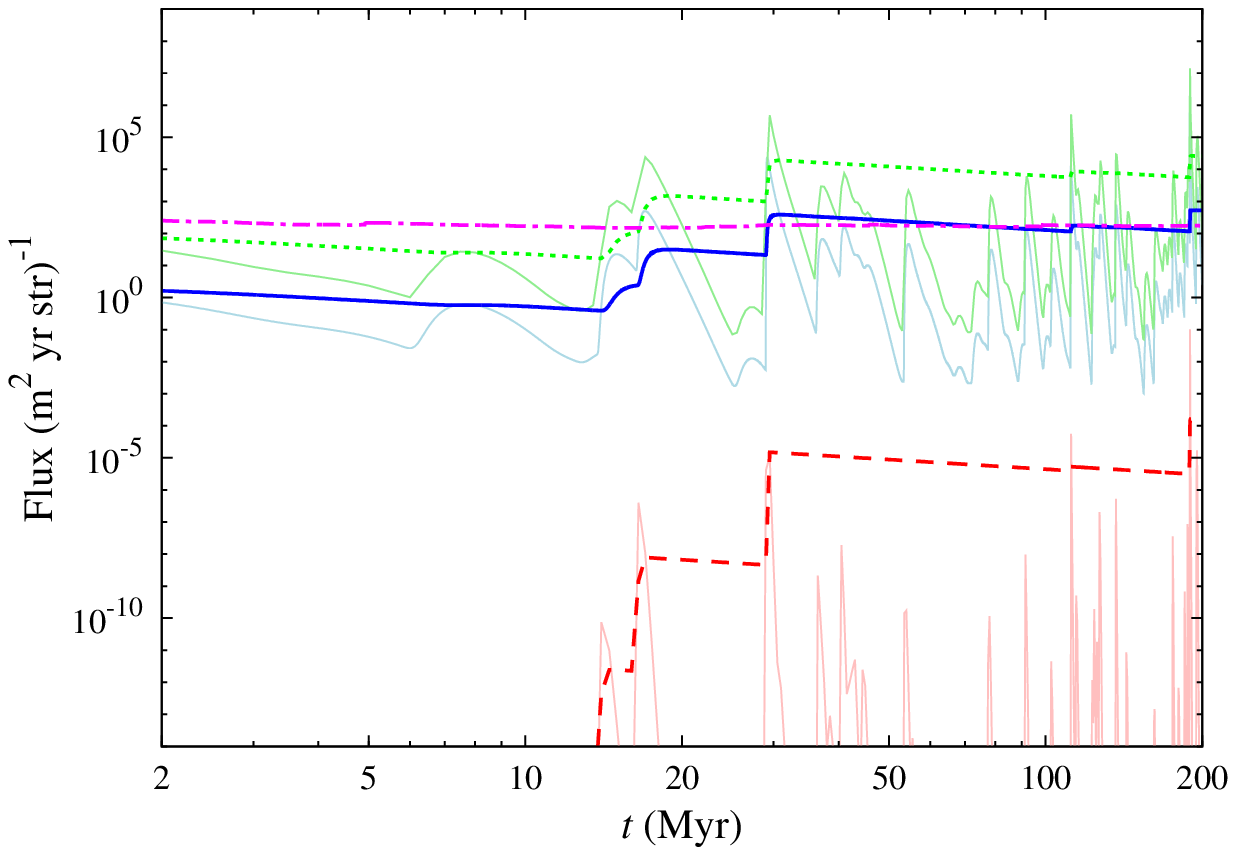}
\includegraphics[width=0.5\textwidth]{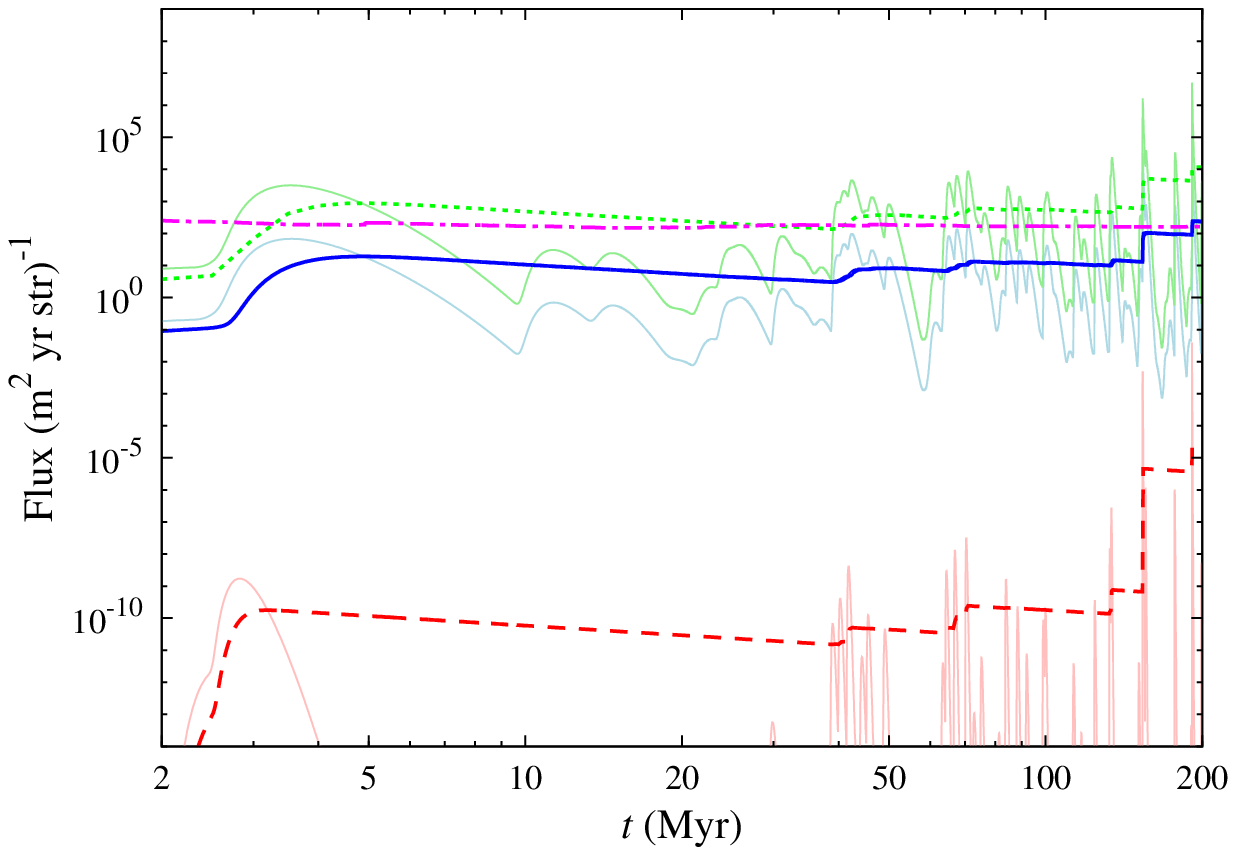}
\includegraphics[width=0.5\textwidth]{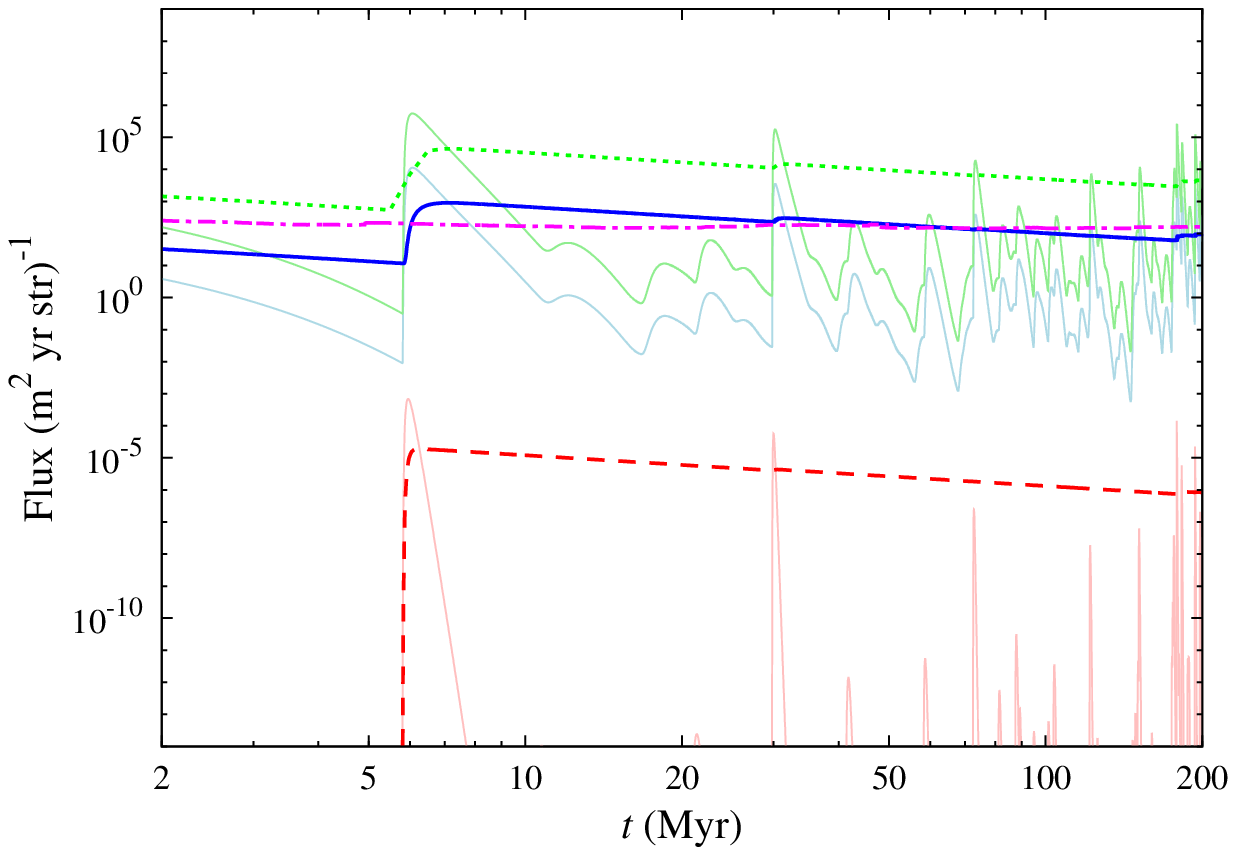}
\caption{ 
UHCR flux averaged over the period from $t=0$ to $t$ for models \red{NSM-RS (blue solid), NSM-MB (red dashed),} NSM-P (green dotted), and SNR (magenta dash-dotted) in the meteorite energy range. 
Thin lines show the instantaneous flux (the same as Figure~\ref{meteo} but a different scale) for models 
\red{NSM-RS (light-blue), NSM-MB (pink), and} NSM-P (light-green). 
Four panels show results with different positions of $\theta = 0, \pi/2, \pi, 3\pi/2$ on a ring with $r = 8$ kpc. 
}\label{cumulative}
\end{figure*}

Meteorite experiments have the possibility to provide further information about NSM-UHCRs because meteorites are exposed to cosmic radiation over millions of years, which is in contrast to the satellite experiments measuring the current flux.
For example, two meteorites with ages of $\sim 70$ Myr and $\sim 200$ Myr were used in the OLIMPIYA experiment. 

In Figure~\ref{cumulative}, we show the  flux averaged over the period  from $t=0$ to $t$ as
\begin{equation}
\frac{1}{t}\int^t_0 F(t) dt , 
\end{equation}
 for the long-term (200 Myr) computation run. 
We show results at four positions of $\theta = 0, \pi/2, \pi, (3/2) \pi$ on a circle of the radius $r =8$ kpc from the MW center.   
The predicted average flux from model NSM-RS is $\sim 10^2 - 10^3  ({\rm m^2\ yr\ str})^{-1}$
 when we use a meteorite with $\sim 200$ Myr. 
 The averaged flux of unshocked NSM ejecta is $10^{-6}$ - $10^{-4} ({\rm m^2\ yr\ str})^{-1}$ and $\sim 10^4 ({\rm m^2\ yr\ str})^{-1}$ for model NSM-MB and NSM-P, respectively. 
The cumulative flux over such a long exposure time is dominated by rare but strong peaks triggered from nearby NSMs. 
At $t = 200$ Myr, \red{the cumulative flux of models NSM-RS is comparable to that of model SNR 
 though the instantaneous flux becomes stronger than model SNR only in a few million years after NSMs in the solar vicinity.   
The averaged flux in model NSM-P overwhelms model SNR for a meteorite with an age of ten million years or more.} 
If the UHCR flux measured by a meteorite with a long age is much higher than the flux obtained by satellite experiments, 
 it indicates a strong contribution from NSM-UHCRs.

This figure also indicates that the average flux is dependent on the age of a meteorite.  
Meteorites with ages below several Myr show very low averaged fluxes of NSM-UHCRs 
 while meteorites with longer exposure times tend to show higher values. 
The fluxes measured by such meteorites are also dependent on the event history in the solar vicinity. 
In the top left panel, 
 a meteorite with an age longer than 17 Myr can detect a significant contribution from NSM-UHCRs, in model NSM-RS, while in the bottom left panel, 
 only meteorites with ages longer than 150 Myr can detect NSM-UHCRs.

The OLIMPIYA experiment using two meteorites with ages of 70 Myr  and 200 Myr yielded abundance ratios of r/Fe consistent with the solar system abundance ratio though the data from meteorites are statistically less significant in the region of $Z<56$. 
If there were a significant difference between the estimated UHCR flux from the two meteorites,
 it would be evidence of contribution from NSMs.  
The authors, unfortunately, do not show results for each satellite.

\citet{Herzog15} estimated the cosmic ray exposure ages of 19 pallasites from the abundances of nuclides produced by cosmic rays. 
The estimated age ranges from 7 Myr to 180 Myr. 
Future experiments using pallasites with various cosmic ray exposure ages have possibility to identify the contribution from NSMs and 
 reconstruct the event history of NSMs in the solar vicinity. 
The improvement in age estimation method is also required   
 since there are still significant discrepancies between ages estimated from different elements. 
For example, the estimated age of the Marjalahti meteorite from $^{36}{\rm Ar}/^{36}{\rm Cl}$ is $185\pm19$ Myr while $^{21}{\rm Ne}$ indicates $43\pm6$ Myr.

\subsection{Composition}

NSM-UHCRs are also different from SNR-UHCRs in the abundance patterns. 
NSM-UHCRs consist entirely of r-process elements while  
 SNR-UHCRs contain both r-process and s-process elements.

Observationally, the very low Pb/Pt ratio in cosmic rays is consistent with NSMs as the dominant UHCR sources, 
 though the volatility biased acceleration model can also explain the paucity of Pb \citep{Meyer97}. 
The detection of the abundance peak at $_{56}$Ba, element of the second peak of s-process, in both satellite and meteorite experiments 
 indicates that the SNR-UHCR flux is not much weaker than the NSM-UHCR flux.

In order to discuss elements apart from the abundance peaks, 
 we have to consider the spallation products. 
Though we do not compute the change of abundance pattern by spallation,
 we can discuss the contribution of the secondary elements from the value of $\nu$ (Eq. (\ref{eq:nu})). 
The flux of primary cosmic rays decreases proportional to $e^{-\nu}$ 
 and the secondary elements are produced as $e^{-\nu}$ decreases. 
We define the spallation fraction, $\langle 1-e^{-\nu} \rangle$, the weighted average of $1-e^{-\nu}$ as
\begin{eqnarray}
&& \langle 1-e^{-\nu} \rangle = \frac{1}{4\pi F(t)} \Biggl[ \int^{E_{\rm up}}_{E_{\rm low}} \sum_i \biggl( \int 1-e^{-\nu(E_0, t-t_i)}  \nonumber  \\
&& Q(E_0)  G(E, E_0, \vec{r}, \vec{r}_i, t, t_i) dE_0 \biggr) \beta(E) dE \Biggr] . 
\end{eqnarray}
 and the cumulative spallation fraction, 
\begin{eqnarray}
\overline{ \langle 1-e^{-\nu} \rangle } =&& \frac{1}{4\pi \int^t_0 F(t) dt} \int^t_0 \Biggl[ \int^{E_{\rm up}}_{E_{\rm low}} \sum_i \biggl( \int 1-e^{-\nu(E_0, t-t_i)}  \nonumber  \\
 Q(E_0) && G(E, E_0, \vec{r}, \vec{r}_i, t, t_i) dE_0 \biggr) \beta(E) dE \Biggr] dt  , 
\end{eqnarray}
for meteorite experiments. 
These values are an index of relative abundances of secondary elements produced by spallation
 neglecting differences in decay rates and energies between the primary and secondary nuclei.

\begin{figure}
\includegraphics[width=\columnwidth]{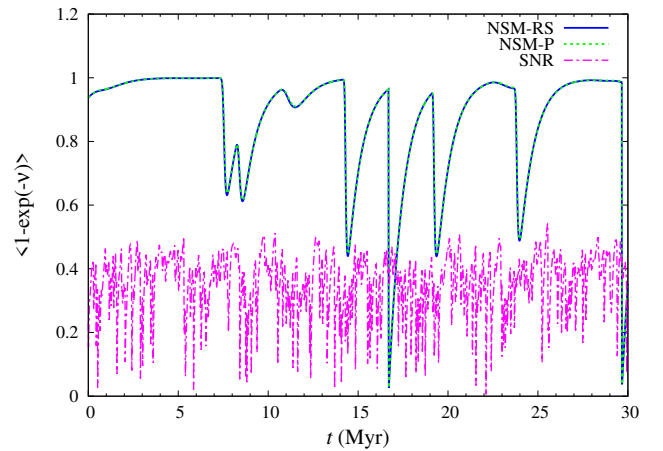}
\caption{ 
Spallation fractions $ \langle 1-e^{-\nu} \rangle$ in the energy range of the UHCRE satellite in models NSM-RS (blue \red{solid}), NSM-P (green dotted), and SNR (magenta dash-dotted). 
The results of NSM-RS and NSM-P are almost overlapped.  
}\label{nuSat}
\end{figure}

Figure~\ref{nuSat} shows
  the spallation fractions at the energy range of the UHCRE satellite in model NSM-RS (blue solid), NSM-P (green dotted) and SNR (magenta dash-dotted). 
The spallation fractions also fluctuate with time. 
When an NSM takes place in the solar vicinity, the spallation fraction drops simultaneously with an increase of the UHCR flux. 
This is because cosmic rays from nearby NSM events can reach the solar system before spallation.

In satellite observations, 
 the composition of NSM-UHCRs can be strongly affected by spallation. 
Most of NSM-UHCRs become secondary products by spallation in 1 Myr after a nearby NSM event. 
However, in model NSM-RS, the predicted NSM-UHCR flux overwhelms the flux of SNR-UHCRs only when an NSM takes place in the solar vicinity. 
When NSM-UHCRs are dominant, the spallation fraction is comparable to that of SNR-UHCRs. 
In NSM-P, secondary elements produced by spallation are expected to be detected by satellites
 at several Myr after an NSM event. 
The spallation fractions of NSM-RS and NSM-P are almost the same since the spallation rate is almost constant above $E_{\rm low, sat}$.

\begin{figure*}
\includegraphics[width=0.5\textwidth]{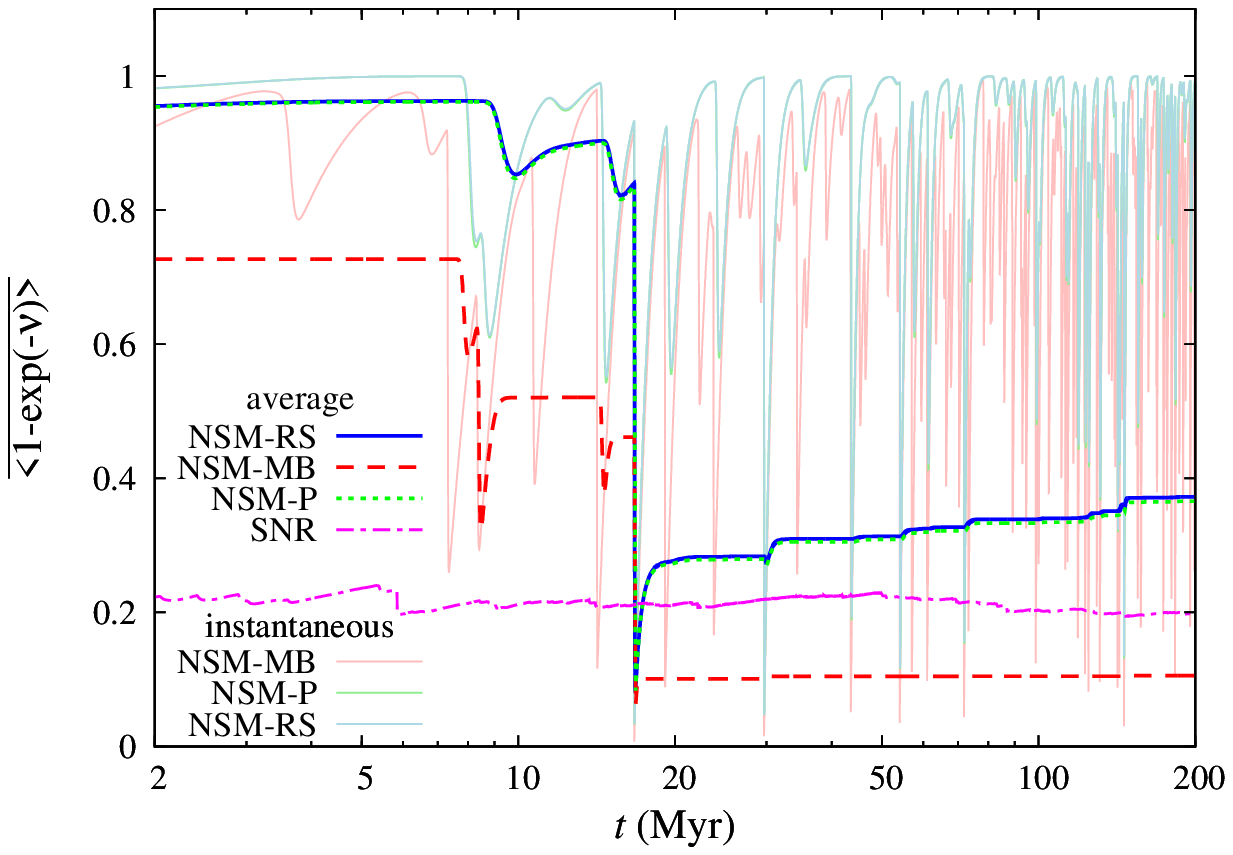}
\includegraphics[width=0.5\textwidth]{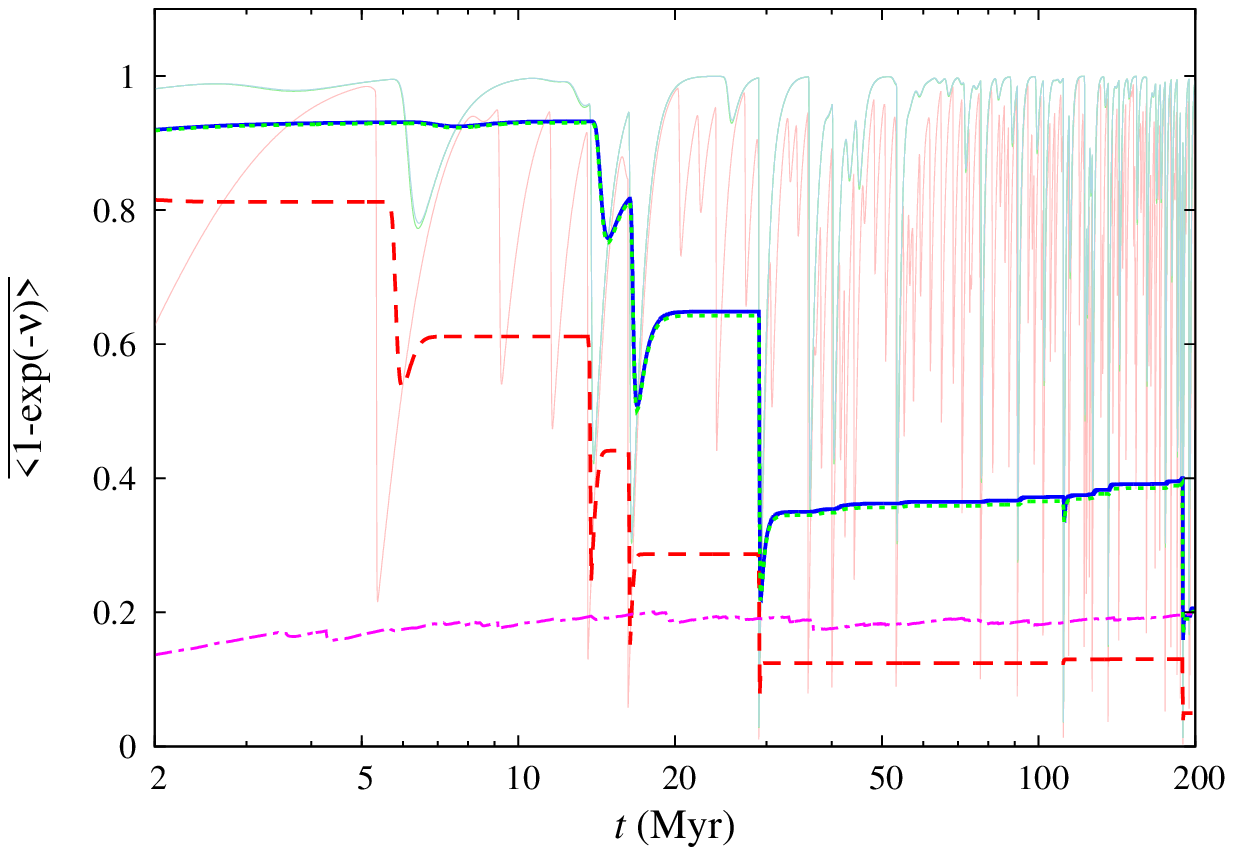}
\includegraphics[width=0.5\textwidth]{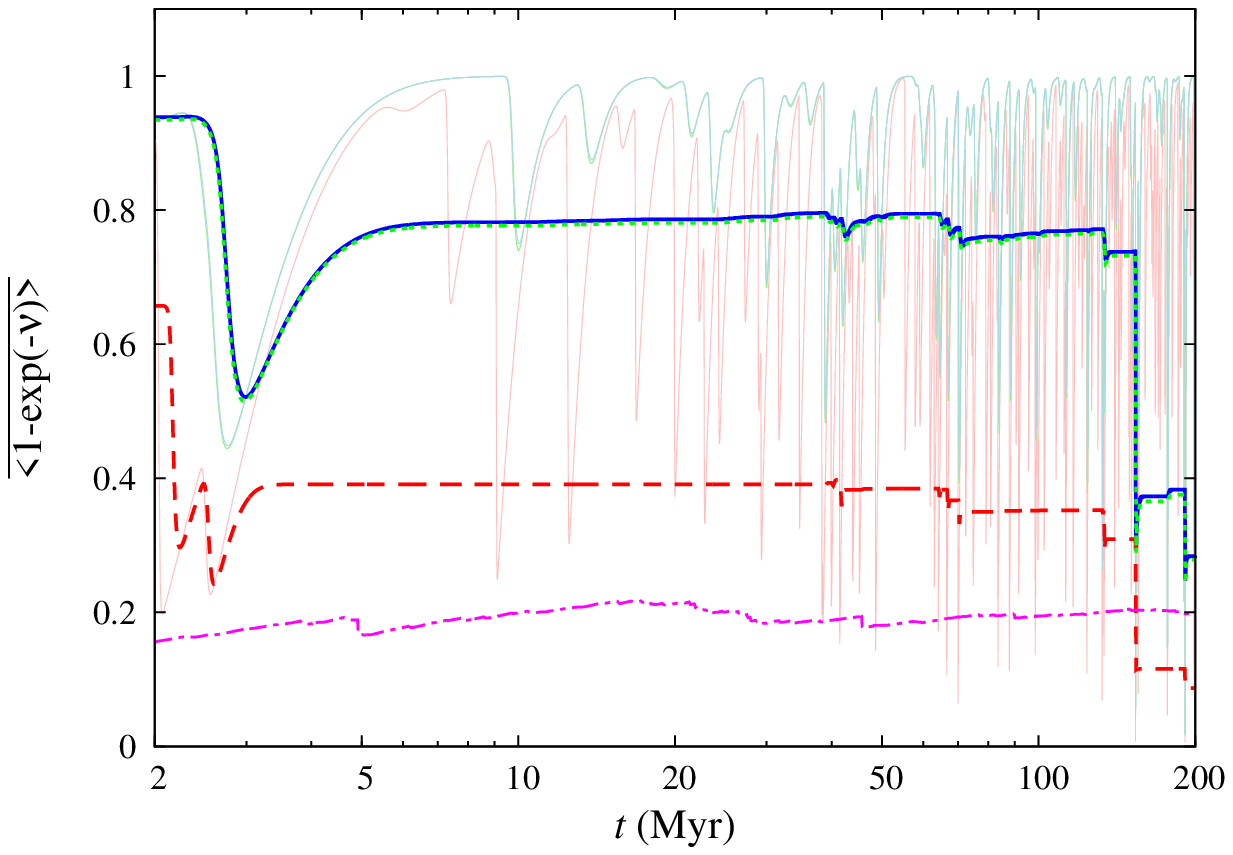}
\includegraphics[width=0.5\textwidth]{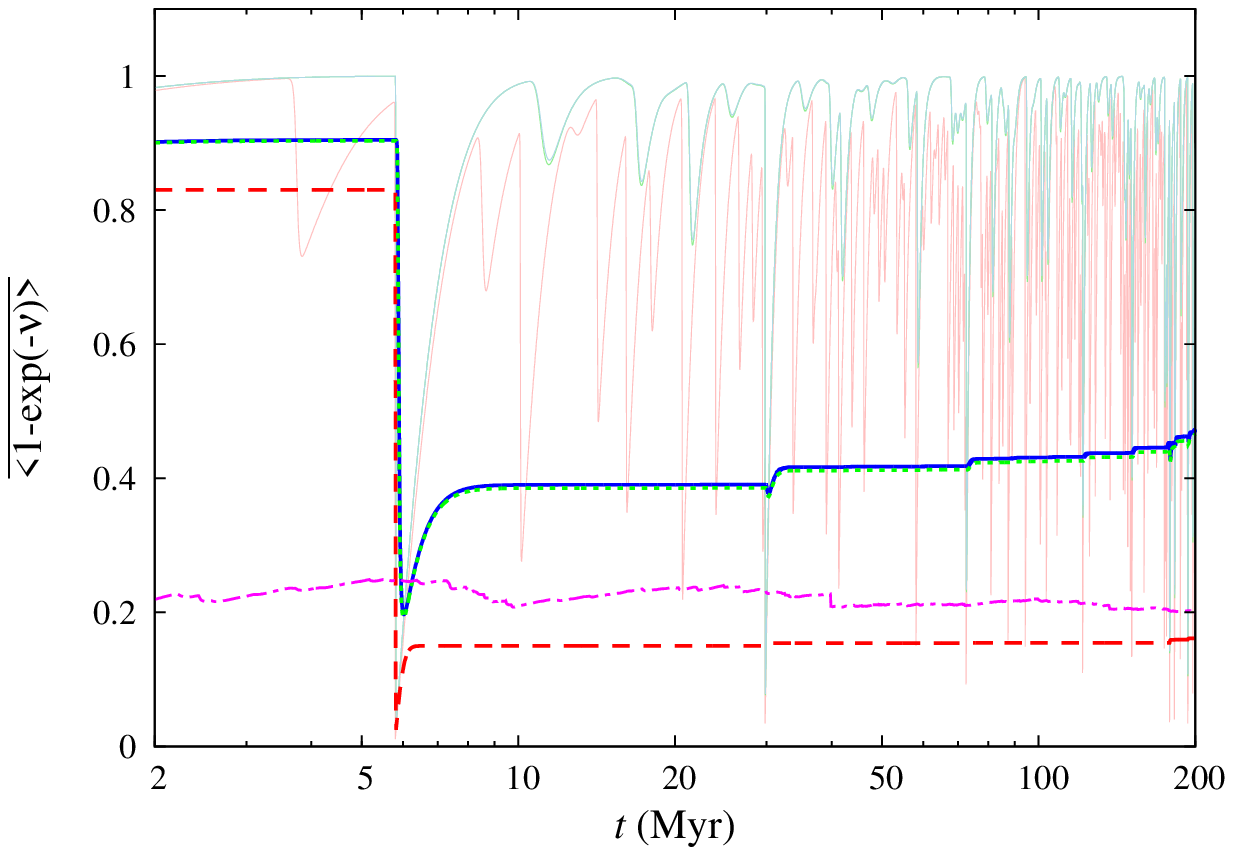}
\caption{ 
 Time averaged spallation fractions $\overline{ \langle 1-e^{-\nu} \rangle }$ in models \red{NSM-RS (blue solid), NSM-MB (red dashed)}, NSM-P (green dotted), and SNR (magenta dash-dotted) in the meteorite energy range.  
 Thin lines show the instantaneous value $ \langle 1-e^{-\nu} \rangle$ in models NSM-RS (light-blue), NSM-MB (pink), and NSM-P (light-green). 
\red{The lines of NSM-RS and NSM-P are almost overlapped.}
}\label{nuMet}
\end{figure*}

Figure~\ref{nuMet} shows the spallation fractions in the meteorite energy range (thin lines)
 and its time average $\overline{ \langle 1-e^{-\nu} \rangle }$ (thick lines).   
\red{
In models NSM-RS and NSM-P, the contribution of spallation products is higher than in SNR-UHCRs, excepting shortly after NSM events in the solar vicinity. 
In particular, in the case of the bottom left panel, most of NSM-UHCRs are the secondary products for the first 150 Myr. 
Because of the long distance from NSMs to the solar system, NSM-UHCRs have experienced collisional fragmentation. 
I}n model NSM-MB, the predicted spallation fractions are smaller than the other models. 
This is because the ionization energy loss timescale is shorter than the fragmentation timescale in the low energy range. 
UHCRs detected by meteorites are dominated by particles that arrive at the solar system before losing the energies due to ionization and before undergoing spallation.

From the observational side, the  abundance pattern in UHCRs measured by the OLIMPIYA experiment exhibits the third peak  lower than those in the solar abundance and measured by the UHCRE satellite, 
 while the abundances of elements between the second and third peaks tend to be higher \citep{Alexeev16}. 
It may indicate a stronger contribution of spallation products in the meteorite experiment than in the satellite experiment,
 while the measurement uncertainty is still very large and there are also similar disagreements between satellite experiments of UHCRE and Ariel-6. 
If the difference between the OLIMPIYA result and the UHCRE result is real, 
 it can be explained by 
\red{contribution of NSM-UHCRs in the meteorite experiment. 
In model NSM-RS or NSM-P, $\overline{ \langle 1-e^{-\nu} \rangle }$ is significantly higher than model SNR. 
In addition, the cumulative flux of NSM-UHCRs is comparable to or larger than that} of SNR-UHCRs, as seen in Figure~\ref{cumulative}. 
Therefore the models predict a high ratio of secondary to primary elements for the meteorite experiment. 
On the other hand, the instantaneous flux of NSM-UHCRs fluctuates and \red{usually smaller than the flux of SNR-UHCRs in model NSM-RS.}  
In such a case, the secondary to primary ratios measured by satellite experiments are dominated by SNR-UHCRs, and smaller than the value measured by meteorites.

\subsection{Energy Spectrum}

\begin{figure}[h]
\includegraphics[width=\columnwidth]{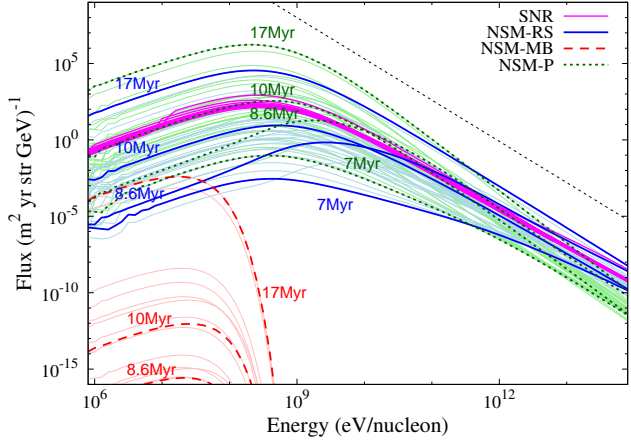}
\caption{Predicted energy spectra of UHCRs.  
Thick blue, red, and green lines are spectra of models NSM-RS, NSM-MB, and NSM-P respectively, at $t= 7, 8.6, 10$ and 17 Myr.
At $t= 7$ Myr, the flux of model NSM-MB is below the plotted range. 
Magenta lines denote spectra of SNR-UHCRs. 
Thin light-blue, pink and light-green lines show spectra at every $10^6$ yrs. 
A dotted black line denotes $p^{-2.7}$. 
}\label{spectra}
\end{figure}

The predicted energy spectrum of model NSM-MB 
 is very different from the shock accelerated UHCR. 
The thick dashed red lines in Figure~\ref{spectra} denote spectra at $t=8.6 \times 10^6, 1.0 \times 10^7$ and $1.7 \times 10^7$ yr for model NSM-MB 
 and the pink lines show the UHCR spectra at every $10^6$yr.

The energy spectrum of model NSM-RS at the solar system can be different from  that of model SNR,  
 though we assume the same initial spectrum.  
In Figure~\ref{spectra}, the blue thick solid lines show the predicted spectra of model NSM-RS at $t=7 \times 10^6, 8.6 \times 10^6, 1.0 \times 10^7$ and $1.7 \times 10^7$ yr. 
At $t=8.6 \times 10^6$ yr, 
 while the predicted flux at \red{$10^{11}$ eV/$A$ is comparable to that from model SNR, 
 the flux below $10^{9}$ eV/$A$ is $\sim 3$ dex or more} smaller. 
This is because of the rigidity dependent diffusion coefficient. 
At this time, the NSM-UHCR flux is growing, as seen in Figure~\ref{satellite} and \ref{meteo},
 by contribution from an NSM occurred at $t=8\times 10^6$yr and 2 kpc from the solar system. 
UHCRs with higher energies reach the solar system earlier because of the larger diffusion coefficients.

Model NSM-P also predicts time dependent spectra, as shown in the green lines in the figure. 
This model shows a steeper slope than SNR-UHCR in the energy range above $\sim1$ TeV/$A$.


\section{Summary and Conclusions}\label{conclusionS}

In this paper, we investigate the contribution of NSMs to UHCRs. 
In particular, we consider the energy loss processes and spallation of UHCRs in the ISM. 
We build a diffusion model of UHCRs with these decay processes and estimate the UHCR flux at the solar system. 
We treat each NSM event discretely since the time span between NSM events are longer than the timescales of the decay processes.

We consider two types of UHCRs from NSMs. 
One assumes that cosmic rays are accelerated at reverse shock in the NSM ejecta (model NSM-RS). 
The other assumes material directly ejected from the NSMs without additional diffusive shock accelerations. 
Since ejecta from NSMs have very large velocities, 
 the ejecta particles which have not experienced reverse shock can be detected as cosmic rays.  
We adopt two model energy spectra for NSM ejecta without shock accelerations, the Maxwell-Boltzmann distribution (model NSM-MB) and the power-law distribution (model NSM-P).

UHCRs are observed using satellite instruments or stony iron meteorites. 
Satellite instruments detect cosmic rays with energies of 1.5 - 10 GeV/$A$ and 
 stony iron meteorites can be used as detectors of UHCRs in the energy range of 2.5 - 300 MeV/$A$. 
Meteorites tell us the cumulative cosmic ray flux over the past millions of years. 
We compute the UHCR flux from NSMs in these energy ranges at the solar system as a function of time, and compare them with the UHCRs accelerated in SNRs. 
We also estimate the amount of secondary elements as spallation products, and show the predicted energy spectra.

The main conclusions are as follows.

\begin{itemize}
\item
Spallation plays an important role in r-process element cosmic rays from NSMs in all the energy range.  
In addition, UHCR nuclei from NSMs at $\lesssim 1$GeV/$A$ significantly lose their energies by ionizing the ISM before they propagate to the solar system. 
These are in contrast with protons from SNRs for which decay processes are negligible excepting at very low energies ($\lesssim 20$MeV/$A$). 
This is because the long propagation timescale due to the very low event rate of NSMs
 in addition to atomic mass dependence of spallation rate and charge dependence of ionization rate.

\item
UHCR flux from NSMs fluctuates with time over many orders of magnitude. 
Model NSM-RS predicts that an NSM event at $\lesssim 1.5$ kpc from solar system would produce the UHCR flux greater than that of SNR-UHCRs, though only several NSM events per hundred million years occur in such the solar vicinity. 
In a few million years from the nearby event, 
spallation decreases the flux by two or three orders of magnitude in the energy range of 1.5 - 10 GeV/$A$. 
In model NSM-P, the predicted flux is 20 - 40 times larger. 
\red{In model NSM-MB, the UHCR flux is greatly reduced by the solar wind and becomes hard to be detected because of the lack of high energy components in the energy spectrum.  }

\item
For the experiments of UHCRs using stony iron meteorites, 
 NSMs are expected to be the dominant source in spite of the short time span when NSM-UHCRs overwhelm SNR-UHCRs, 
 \red{if NSM-UHCRs have high energy component. }
This is because that very strong irradiation from rare events in the solar vicinity dominates the cumulative UHCR flux of meteorite experiments with very long exposure times. 
The time averaged flux of NSM-UHCRs is expected to be comparable to or larger than that of SNR-UHCRs.

\item
Though the current observational data are not enough to establish or refute the contribution from NSM-UHCRs,
future experiments using multiple stony-iron meteorites with different ages can reveal the contribution. 
The cumulative nature of meteorite observations can record a strong cosmic ray irradiation from nearest NSM events if they happened while the meteorite is exposed to cosmic rays.
If the time averaged flux measured by meteorites with different ages show different value, 
 it will be a smoking gun of the NSM contribution to UHCRs. 
In addition, we have possibility to reconstruct the event history of NSMs in the solar vicinity.

\item
Meteorite experiments with a long exposure time are likely to detect NSM-UHCRs rather than SNR-UHCRs while satellite experiments tend to detect SNR-UHCRs. 
If this is the case, the abundance pattern of UHCRs measured by these different experiments can be different 
 because NSM-UHCRs suffer from spallation and fluctuate the abundance of the spallation products depending on the distances to NSM sites. 
If the disagreement between abundance patterns obtained from the OLIMPIYA experiments using meteorites and the UHCRE experiment using a satellite is confirmed,
 it would support this NSM scenario, though there are still very large uncertainties in observed abundances.

\item
In model NSM-MB, the expected energy spectrum is significantly different from that of SNR-UHCRs. 
The predicted spectrum 
 shows exponential cutoff at energies around $100 {\rm MeV}/A$.  
The energy spectrum of the shocked UHCRs from NSMs can also be different from SNR-UHCRs,  
 even when we assume the same initial energy spectrum. 
At the growing phase of UHCR flux from a nearby NSM, energy spectra at 1 - 100 GeV/$A$ become shallower 
 because of the energy dependence of the diffusion coefficient.

\end{itemize}

We compute the total flux of r-process elements and estimate the fraction of spallation products by simple model, but do not investigate the detailed abundance patterns. 
It should be investigated in future works considering acceleration bias and change of composition by spallation.

Though we assume the uniform density of ISM for simplicity, 
 local density distribution can affect the propagation of UHCRs \citep{Combet05} in addition to the difference of density between disk and halo.  
In this paper, NSMs are assumed to take place on the Galactic plane.  
The spatial distribution of observed binary pulsars in the MW supports this assumption 
 while observations of SGRBs indicate they are also distributed around the outer halo region.

We still have poor observational knowledge about both UHCRs and NSMs. 
There are still significant uncertainties on the absolute flux and the abundance pattern of UHCRs. 
There are few observational data for the cosmic ray energy spectrum of elements heavier than iron. 
We still have no observational sample of confirmed NSMs, though many SGRBs and a few kilonova candidates are detected. 
Future observations of the abundance pattern and spectrum of UHCRs and their temporal evolutions 
 will be useful probes to understand the nature of NSM ejecta and origin of r-process elements
 in combination of detailed studies of propagation of UHCRs. 

\red{
\begin{acknowledgements}
This work was supported by the JSPS KAKENHI Grant Number 16H02168 and 16H06341. 
\end{acknowledgements}
}


\end{document}